\documentclass[amsmath,amssymb,aps,physrev,reprint,superscriptaddress,footnoteinbib%
,longbibliography
]{revtex4-2} 
\usepackage{amsmath,amssymb}
\usepackage[english]{babel}
\usepackage{ bbold }
\usepackage{upgreek}
\usepackage{bm}
\usepackage{dsfont}
\usepackage{graphicx}
\usepackage{comment}
\usepackage{braket}
\usepackage{colortbl}
\usepackage{blindtext}
\usepackage[hidelinks]{hyperref}
\usepackage{lmodern}
\usepackage{lettrine}
\usepackage{xcolor}
\usepackage[caption=false]{subfig}

\usepackage{tikz}

\newcommand*\diff{\mathop{}\!\mathrm{d}}
\newcommand*\Diff[1]{\mathop{}\!\mathrm{d^#1}}
\newcommand{\tr}{\text{tr}}

\newcommand{\proj}[1]{ \ket{#1} \kern-3pt \bra{#1} }
\newcommand{\TR}{\text{Tr}} 

\begin{document}

\title{Decoherence of dielectric particles by thermal emission}
	
\author{Jonas Sch\"afer}
\affiliation{Faculty of Physics, University of Duisburg-Essen, Lotharstra\ss e 1, 47048 Duisburg, Germany}

\author{Benjamin A. Stickler}
\affiliation{Institute for Complex Quantum Systems,
Ulm University - Albert-Einstein-Allee 11, D-89069 Ulm, Germany}
\affiliation{Faculty of Physics, University of Duisburg-Essen, Lotharstra\ss e 1, 47048 Duisburg, Germany}

\author{Klaus Hornberger}
\affiliation{Faculty of Physics, University of Duisburg-Essen, Lotharstra\ss e 1, 47048 Duisburg, Germany}

\begin{abstract}
Levitated nanoparticles are a promising platform  for sensing applications and for macroscopic quantum experiments. 
While the nanoparticles' motional temperatures can be reduced to near absolute zero, their uncontrolled  internal degrees of freedom remain much hotter, inevitably leading to the emission of heat radiation.
The decoherence and motional heating caused by this thermal emission process  is still poorly understood beyond the case of the center-of-mass motion of point particles.
Here, we present the master equation describing the 
impact of heat radiation on the motional quantum state  
of arbitrarily sized and shaped dielectric rigid rotors.
It predicts the localization of spatio-orientational superpositions only based on the bulk material properties and the particle geometry. 
A counter-intuitive and experimentally relevant implication of the presented theory is that orientational superpositions of optically isotropic bodies are not protected by their symmetry, even in the small-particle limit.
\end{abstract}

\maketitle

\section{Introduction}
Nanoparticles levitated in high vacuum allow for the precise sensing of forces and torques, and will enable macroscopic quantum experiments, due to their exquisite insulation from  environmental disturbances \cite{Gonzalez-Ballestero2021}.  As a recent milestone, optically trapped nanoparticles have been driven into the quantum mechanical ground state of their centre of mass motion, using the coherent scattering of tweezer light in into a high-finesse cavity \cite{Delic2020, Ranfagni2022, Piotrowski2023} or feedback cooling schemes based on detecting  the scattered light \cite{Magrini2021, Tebbenjohanns2021, Kamba2022}.

The internal degrees of freedom are not cooled in these experiments. At best, they remain in thermal equilibrium with the environment, but   much higher internal temperatures are expected in presence of optical fields due to  unavoidable photon absorption. Consequently, the emitted thermal radiation
is considered one of the major obstacles 
for macroscopic quantum superposition tests with spatially delocalized particles
\cite{Hackermueller2004, Chang2009, RomeroIsart2011a, Bateman2014, Gasbarri2021Jul, Neumeier2024, RodaLlordes2024, Steiner2024May}.

Recently, interest has also focused on experiments beyond the small-particle regime; for instance, the Mie resonances arising in 
optically levitated  microscale spheres have been proposed as an additional 
means of manipulation \cite{Lepeshov2023, Maurer2023Sep}. 
Moreover, control over the particles' orientation is expected to enter the quantum regime in the near future \cite{Stickler2016,Seberson2019,Bang2020, Schaefer2021, Rudolph2021, Laan2021, Pontin2023, Kamba2023, Gao2024}, opening up another avenue for quantum experiments based on the intrinsic non-linearity of rotations \cite{Stickler2018, Ma2020, Stickler2021}.
In view of these developments, it is important to quantitatively predict
how the thermal radiation emitted by extended dielectric bodies affects the coherence of their ro-translational quantum dynamics.

In this article, we present the quantum  master equation describing the
impact of thermal photon emission onto the external degrees of freedom of an arbitrarily shaped and sized dielectric rigid body. 
The emitted radiation, which  is sourced from thermally driven polarization currents within the material, contains information on the particle position and orientation.
The ensuing decoherence dynamics are fully characterized by the geometry of the particle and the complex bulk permittivity of its material, while 
internal photon scattering is accounted for in the light-matter interaction to all orders.

To introduce concepts and notation, we first discuss the  master equation  for the rotational motion of particles which are small compared to the thermal wave length. One of the remarkable effects predicted is the decay of orientational superpositions involving states that are optically indistinguishable; even a perfectly homogeneous and isotropic dielectric ball is thus not protected by its symmetry. A microscopic derivation based on the standard Born-Markov approximation
confirms
this effect as originating from the vector character of the fluctuating polarization current.

We then introduce the general master equation for the center-of-mass and rotational motion of arbitrarily sized particles, which can be obtained by inferring the impact of photon emission events due to thermal polarization fluctuations from the backaction associated with scalar particles leaking from metastable binding sites.
The small particle limit reproduces the microscopically derived master equation exactly, while for large particles the decoherence rate exhibits a volume-to-surface transition.
We also discuss the implications of the presented theory for future quantum experiments with submicron particles by calculating the shape-dependent motional heating rates of silica particles for various temperatures.

\begin{figure*}
	\centering
        \subfloat{
\begin{tikzpicture}
    \node[anchor=south west,inner sep=0] (image) at (0,0)
    {\includegraphics[width=0.64\linewidth]{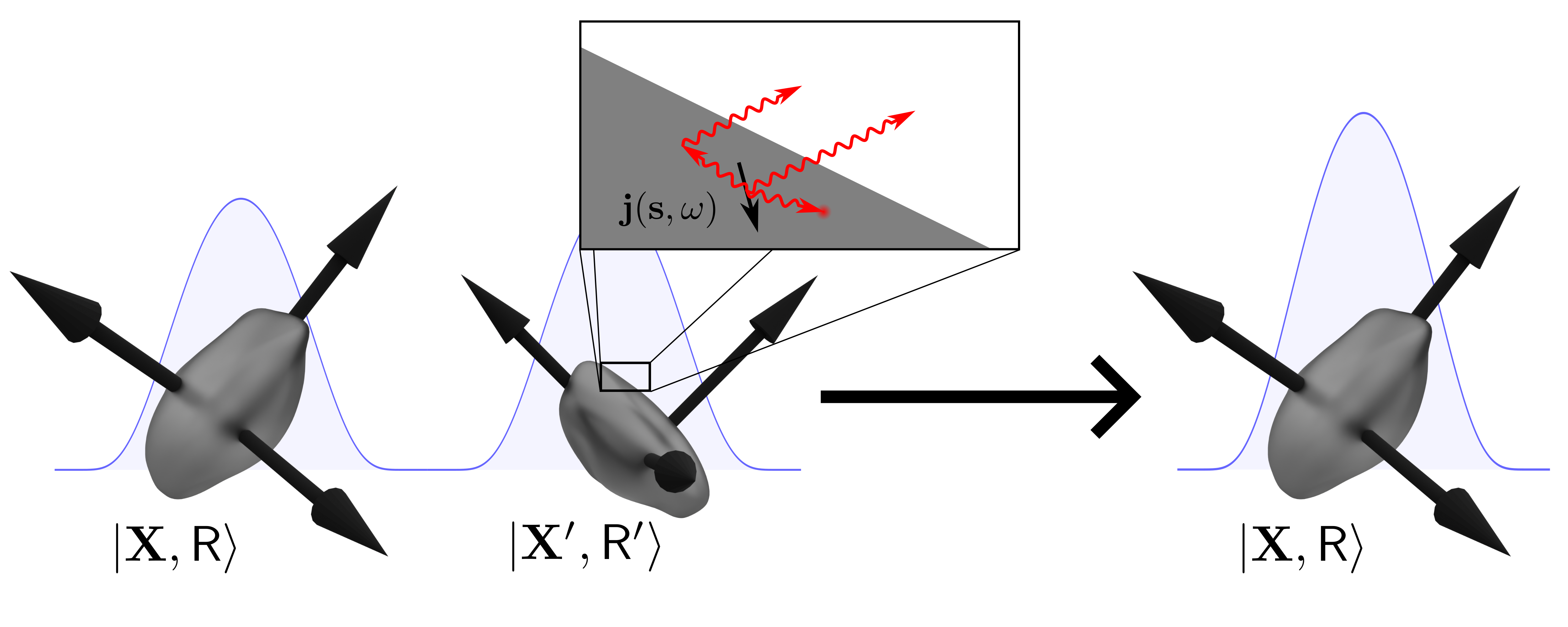}};
    \begin{scope}[x={(image.south east)},y={(image.north west)}]
        \node[] at (0.04,0.96) {\textbf{(a)}};
    \end{scope}
\end{tikzpicture}
        }
        \subfloat{
\begin{tikzpicture}
    \node[anchor=south west,inner sep=0] (image) at (0,0) {\includegraphics[width=0.36\linewidth]{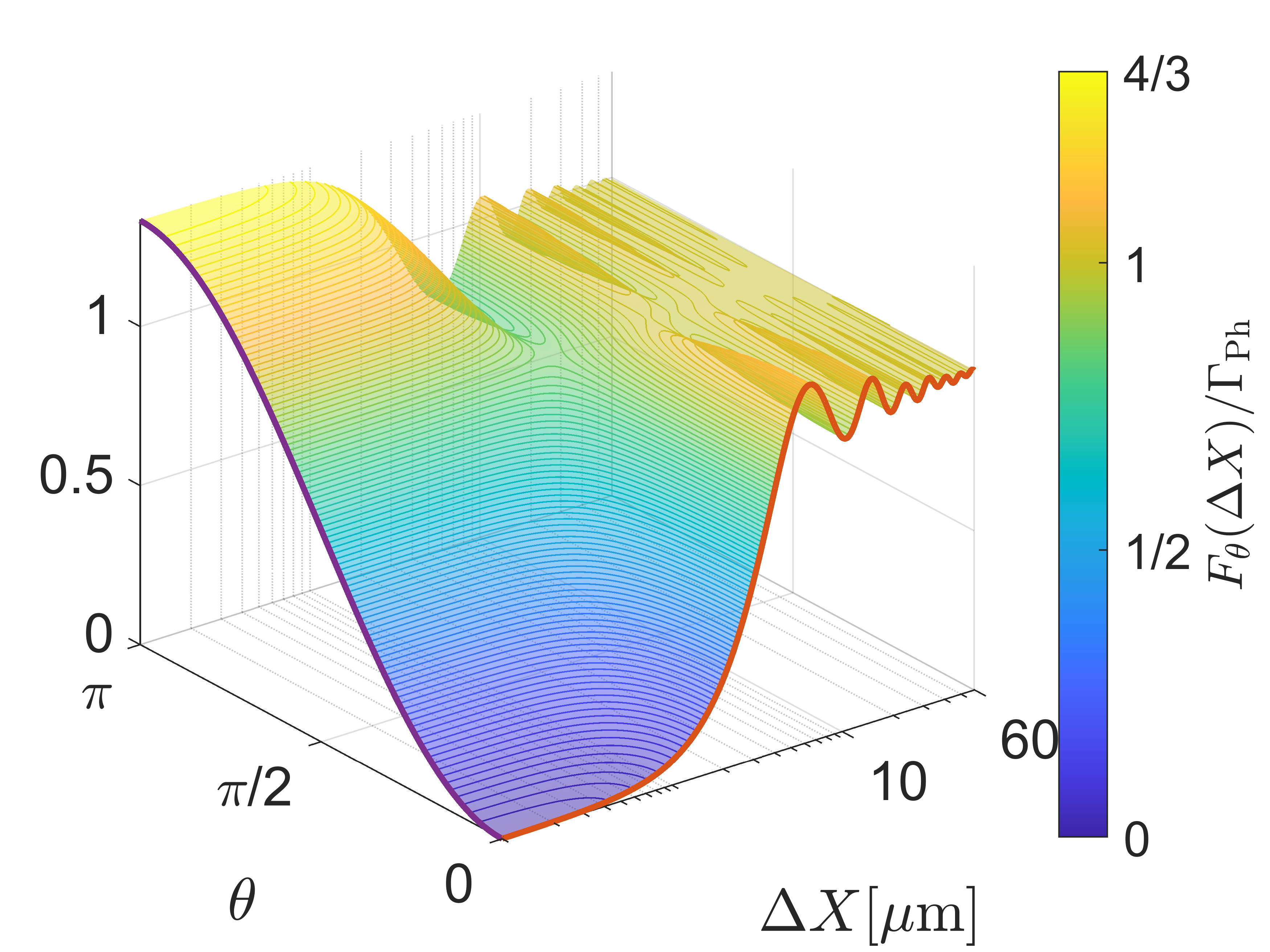}};
    \begin{scope}[x={(image.south east)},y={(image.north west)}]
        \node[] at (0.04,0.906) {\textbf{(b)}};
    \end{scope}
\end{tikzpicture}
        }
    \caption{(a) 
    A dielectric particle  delocalized in position $\mathbf{X}$ and orientation $\mathsf{R}$ emits photons sourced from thermally driven polarization currents $\mathbf{j} (\mathbf{s}, \omega)$.
    This gradually turns the initial quantum superposition into a mixture of states with definite position and orientation, as described by the  Lindblad master equation~(\ref{eq:MEGen}).
 (b) Decoherence rate (\ref{eq:SPDecRaFull}) for a spherical SiO$_{2}$ nanoparticle with mass $10^9\,$amu ($113$\,nm diameter) at $1000$\,K for a superposition generated by a rotation around and a translation along a common vector, as a function of the angle of rotation
$\theta({\mathsf{R},\mathsf{R}'}) $ and the separation $\Delta X$; the total photoemission rate is $\Gamma_\mathrm{Ph}=889\,$MHz.
 \label{fig:Main}
 }
\end{figure*}

\section{Motional decoherence by heat radiation}

The impact of thermal emission on the combined translational and rotational state $\rho$ of a dielectric rigid body can be described by a Markovian master equation of the form $\partial_t\rho=-i[\hat{H},\rho]/\hbar+\mathcal{D}\rho$,  where external forces and torques are included in the Hamiltonian $\hat{H}$. We focus in the following on the incoherent part $\mathcal{D}\rho$, which
accounts for ro-translational decoherence and diffusion.
The main assumptions
underlying our derivation are: (1) the optical properties of the nanoparticle are fully characterized by its dielectric function and geometric shape, (2) its heat capacity is sufficiently large such that the internal state remains in quasi-equilibrium, and (3) the thermal radiation is emitted into the free vacuum field.

We denote the center-of-mass position of the nanoparticle by the vector $\mathbf{X}$ and its orientation by the rotation tensor $\mathsf{R}$; the latter describes how a body-fixed vector transforms when the particle is rotated from a reference orientation to its actual orientation.
Quantum mechanically, these quantities are promoted to operators $\hat{\mathbf{X}}$ and $\hat{\mathsf{R}}$ with eigenkets $\ket{\mathbf{X},\mathsf{R}}$. We find that $\mathcal{D}\rho$ acts as a multiplication in this basis,
\begin{align}\label{eq:Drho}
    \braket{\mathbf{X}, \mathsf{R} | \mathcal{D} \rho | \mathbf{X}' , \mathsf{R} '}
    ={}&
    - F_{\mathsf{R}, \mathsf{R}'}
    (\mathbf{X}- \mathbf{X}') \braket{\mathbf{X}, \mathsf{R} | \rho | \mathbf{X}' , \mathsf{R} '}
    .
\end{align}
An explicit form of  ${F}_{\mathsf{R}, \mathsf{R}'} (\Delta\mathbf{X})$ is given in (\ref{eq:Ftilde}).
Its real part defines the spatio-orientational localization rate, since it is non-negative and vanishes for diagonal arguments thus describing an exponential decay of the coherences, see Fig.~\ref{fig:Main}(a).

To establish a physical picture of the decoherence process, we start with the reduced dynamics of the orientation state for particles which are small compared to the emitted wave lengths.
The full expression for $\mathcal{D}\rho$ is discussed afterwards, see Eq.~(\ref{eq:MEGen}).

\subsection{Orientational decoherence in the small-particle limit}
The light-matter interaction of a dielectric particle much smaller than all relevant optical wavelengths 
can be characterized by a symmetric polarizability tensor $\upalpha (\omega)$,
whose imaginary part $\upalpha'' \equiv (\upalpha - \upalpha^\dagger)/2 i$ is positive and describes electromagnetic absorption.
In thermal equilibrium (and assuming detailed balance) it thus determines the spectrum of emitted thermal radiation \cite{Bohren1983}, suggesting that decoherence is also governed by this quantity.

Indeed, we find that in the small-particle limit the rate in Eq.~(\ref{eq:Drho}) reduces to the pure orientational localization rate
\begin{align}\label{eq:SPDecRaOr}
		F_{\mathsf{R},\mathsf{R}'} ( 0 )
		={}&
		\int\limits_{0}^{\infty} 
  \dfrac{\diff \omega \,\omega^3}{3 \pi^2 c^3 \epsilon_0} \overline{n} (\omega) \,
		\TR \Big[
  \upalpha'' (\omega)
		\left( \mathds{1} -
  \mathsf{R}^{T} 		\mathsf{R}' 
  \right) \Big],
\end{align}
where $\overline{n} (\omega)=[\exp(\hbar\omega /k_\mathrm{B}T)-1]^{-1}$ denotes the Bose-Einstein occupation and  $\TR [{\cdot}]$ is the tensorial trace (as opposed to the trace $\tr [\cdot]$  over the motional Hilbert space).
Note that the decoherence rate depends  on  $\mathsf{R}$, $\mathsf{R}'$ only through the rotation $\mathsf{R}^{T} \mathsf{R}'$ relating the orientation $\ket{\mathsf{R}}$ to $\ket{\mathsf{R'}}$. The polarizability tensor refers to the reference orientation. 
The non-negativity of (\ref{eq:SPDecRaOr}) follows from the symmetry of $\alpha''$ and 
the orthogonality of the rotation tensors.

Equation (\ref{eq:SPDecRaOr}) admits a clear interpretation of the  decoherence process since the part
involving the identity tensor amounts to the total photon emission rate
$\Gamma_{\text{Ph}} = \int_0^\infty \diff \omega \, \gamma_{\text{Ph}} (\omega)$, as determined by the spectral rate $ \gamma_{\text{Ph}} (\omega) = \omega^3 \overline{n} (\omega) \TR [\upalpha'' (\omega)] / 3\pi^2 c^3 \epsilon_0 $.
The loss of coherence between $\ket{\mathsf{R}}$ and $\ket{\mathsf{R}'}$
is thus given  by the rate of photon emissions weighted by the extent to which each photon can distinguish between these particle orientations.
The associated distance measure in the orientation manifold is induced by the polarization tensor.

The simplest case is a particle with an isotropic polarizability $\upalpha =\alpha \mathds{1}$. The distance measure then simplifies to a function of to the angle of rotation
$\cos \theta({\mathsf{R},\mathsf{R}'}) = ( \TR [\mathsf{R}^{T} \mathsf{R}'] - 1)/2 $ between $\ket{\mathsf{R}}$ and $\ket{\mathsf{R}'}$ and the decoherence rate  (\ref{eq:SPDecRaOr})  reduces to
\begin{align}
F_{\mathsf{R},\mathsf{R}'} (0) ={}
    \dfrac{2}{3}
    \Gamma_{\text{Ph}} 	[1 - \cos
    \theta({\mathsf{R},\mathsf{R}'})]
    .
\end{align}
Remarkably, in the small-particle limit the orientational decoherence rate is thus proportional to the number of photons emitted, independent of their wavelengths.

From a conceptual point of view, it may seem surprising that a particle with isotropic polarizability should give rise to orientational decoherence at all, since its electromagnetic response is the same for all orientations. Indeed, the decoherence induced by the scattering of photons vanishes in this limit \cite{Stickler2016, Papendell2017, Pedernales2020,Seberson2020,Laan2021}.
To clarify 
why the emission process breaks the underlying symmetry, we next confirm Eq.~(\ref{eq:SPDecRaOr}) using the standard Born-Markov methods. 

In the small-particle limit, the dielectric medium can be modeled as a set of independent dipoles oscillating at non-degenerate frequencies $\omega_j$. They are represented by  bosonic mode operators $a_j$ and 
(body-fixed) transition matrix elements $\mathbf{d}_j$.
The interaction with the free electric field $  \mathbf{E} (\mathbf{r})$ at a fixed particle position $\mathbf{X}$ reads
$ \hat{H}_{\text{int}}={}
		- \sum_{j}
  \mathbf{E} (\mathbf{X}) \cdot
    \hat{\mathsf{R}}
		\left[ \mathbf{d}_{j} a_j +
        \mathbf{d}_j^* a_j^\dagger
        \right]
$.
We trace out the electromagnetic vacuum in the standard Born-Markov and rotating wave approximation \cite{HeinzPeterBreuer2007}, while neglecting the free rotation dynamics of the particle (``sudden approximation''). 
The polarizability can now be determined by considering the linear response of the total dipole moment to an additional classical external drive of frequency $\omega$. In  reference orientation this yields the polarizability tensor $\upalpha (\omega) = \sum_j [\omega_j - \omega - i \gamma_j]^{-1} \mathbf{d}_j \otimes \mathbf{d}_j^*/\hbar$ with damping rates $\gamma_j = |\mathbf{d}_j|^2 \omega_j^3/6 \pi \hbar c^3 \epsilon_0$.
To obtain the master equation for the rotational dynamics, we also trace out the internal modes in another Born-Markov approximation, assuming them to be in a thermal state at temperature $T$. In the weak coupling limit $\gamma_j\to 0$ the imaginary part of the polarizability can then be identified in the resulting master equation with the decoherence rate given in Eq.~(\ref{eq:SPDecRaOr}). We note that the center-of-mass dynamics can be included by promoting the position dependence in the interaction Hamiltonian to an operator.

From this microscopic point of view, it can now be understood why orientational superpositions decohere even for optically isotropic small particles. Photon emission entangles the orientational state not only  with the electromagnetic field  but also  with each of the internal oscillators.
By monitoring the internal state to register from which dipole an emission took place, and correlating this with the outgoing dipole radiation pattern, one could thus in principle learn about the particle orientation even for an isotropic distribution of dipoles.
In case of elastic scattering of  electromagnetic radiation,  in contrast,
the internal state remains unchanged and thus separable.
This implies that
only the collective response matters
since one cannot
use the internal state to
determine
at which dipole the scattering took place.
Hence, no information about the orientation state is contained in the outgoing field if this response is isotropic.
From the perspective of macroscopic electrodynamics used in this article, it is the vectorial nature of the thermally driven polarization current density that breaks the isotropy of the radiation sourced from individual fluctuation events.
This is a striking example for a situation where the naively expected symmetry of the object does not imply a corresponding decoherence-free subspace.

In analogy to center-of-mass decoherence giving rise to momentum diffusion, one expects that orientational decoherence is accompanied by angular momentum diffusion.
Inserting the spectral decomposition of 
$\upalpha''$ into Eq.~(\ref{eq:SPDecRaOr}) one obtains a sum of localization rates,
the form of which  was studied in \cite{Papendell2017} (labeled ``type 1''). 
The results of \cite{Papendell2017} then imply  that Eq.~(\ref{eq:SPDecRaOr}) indeed leaves the average angular momentum unchanged,
$ \partial_t\langle\hat{\mathbf{J}}\rangle=0$, while
its second moments evolve according to
\begin{align}\label{eq:JJ}
		\partial_t \braket{\hat{\mathbf{J}} \otimes \hat{\mathbf{J}} } ={}&
\int\limits_0^\infty \diff \omega \, \dfrac{\hbar^2  \omega^3}{ 3\pi^2 c^3 \epsilon_0} \overline{n} (\omega)
\nonumber\\&\times
\left(
\TR [\upalpha'' (\omega)]\mathds{1}
- \braket{ \hat{\mathsf{R}} \upalpha'' (\omega) \hat{\mathsf{R}}^T }
\right)
  .
\end{align}
For particles with isotropic polarizability the diffusion is independent of the orientation state,
$\braket{\hat{\mathbf{J}} \otimes \hat{\mathbf{J}} } = 2 \hbar^2 \Gamma_\text{Ph} \mathds{1}/3$.

The diffusion (\ref{eq:JJ}) determines 
the growth rate of rotational energy associated with the $i$-th principal axis as
\begin{equation}\label{eq:PHPPRot}
    h_{\mathrm{rot}}^i ={}
    \dfrac{\hbar^2}{2 I_i}
    \int\limits_0^\infty \diff \omega \,
    \dfrac{\omega^3}{3 \pi^2 c^3 \epsilon_0} \overline{n} (\omega)
    [\TR \upalpha'' (\omega) - \alpha_i'' (\omega)],
\end{equation}
with $I_i$ the moment of inertia and $\alpha_i''$ the imaginary part of the corresponding eigenvalue of the polarizability tensor.
(Here we assume the  polarizability to be diagonal with respect to the principal axes, as holds for homogeneous ellipsoids.)

\begin{figure*}
	\centering
        \subfloat{
 \begin{tikzpicture}
    \node[anchor=south west,inner sep=0] (image) at (0,0) {\includegraphics[width=0.5\linewidth]{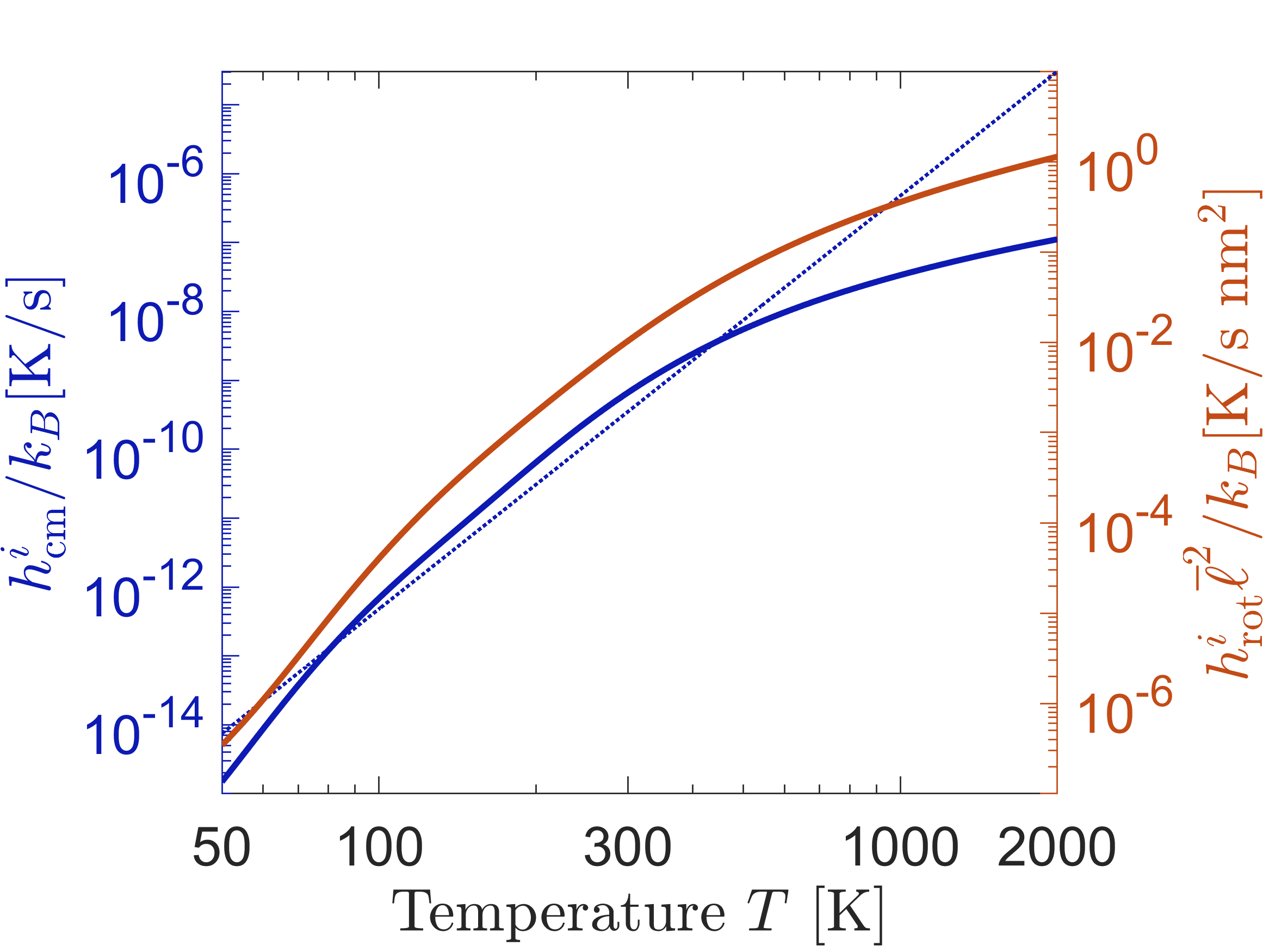}};
    \begin{scope}[x={(image.south east)},y={(image.north west)}]
        \node[] at (0.04,0.98) {\textbf{(a)}};
    \node[anchor=south west,inner sep=0] (image) at (0.485,0.195) 
    {\includegraphics[width=0.16\linewidth]{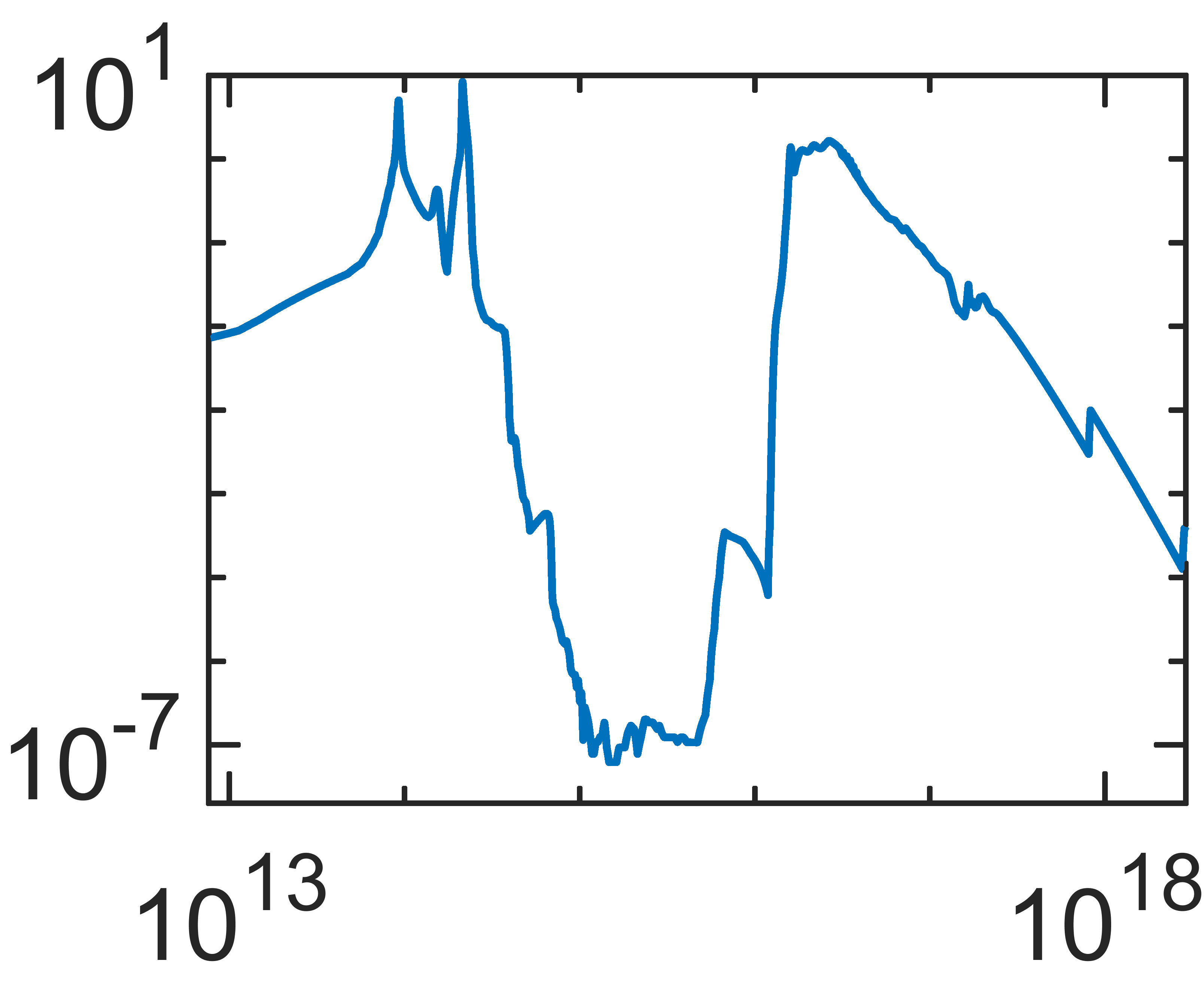}};
    \node[] at (0.67,0.22) {\small $\omega$[1/s]};
    \node[rotate=90] at (0.51,0.4) {\small $\alpha_j''/V \epsilon_0$};
    \end{scope}
\end{tikzpicture}
        }
        \subfloat{
\begin{tikzpicture}
    \node[anchor=south west,inner sep=0] (image) at (0,0) {\includegraphics[width=0.5\linewidth]{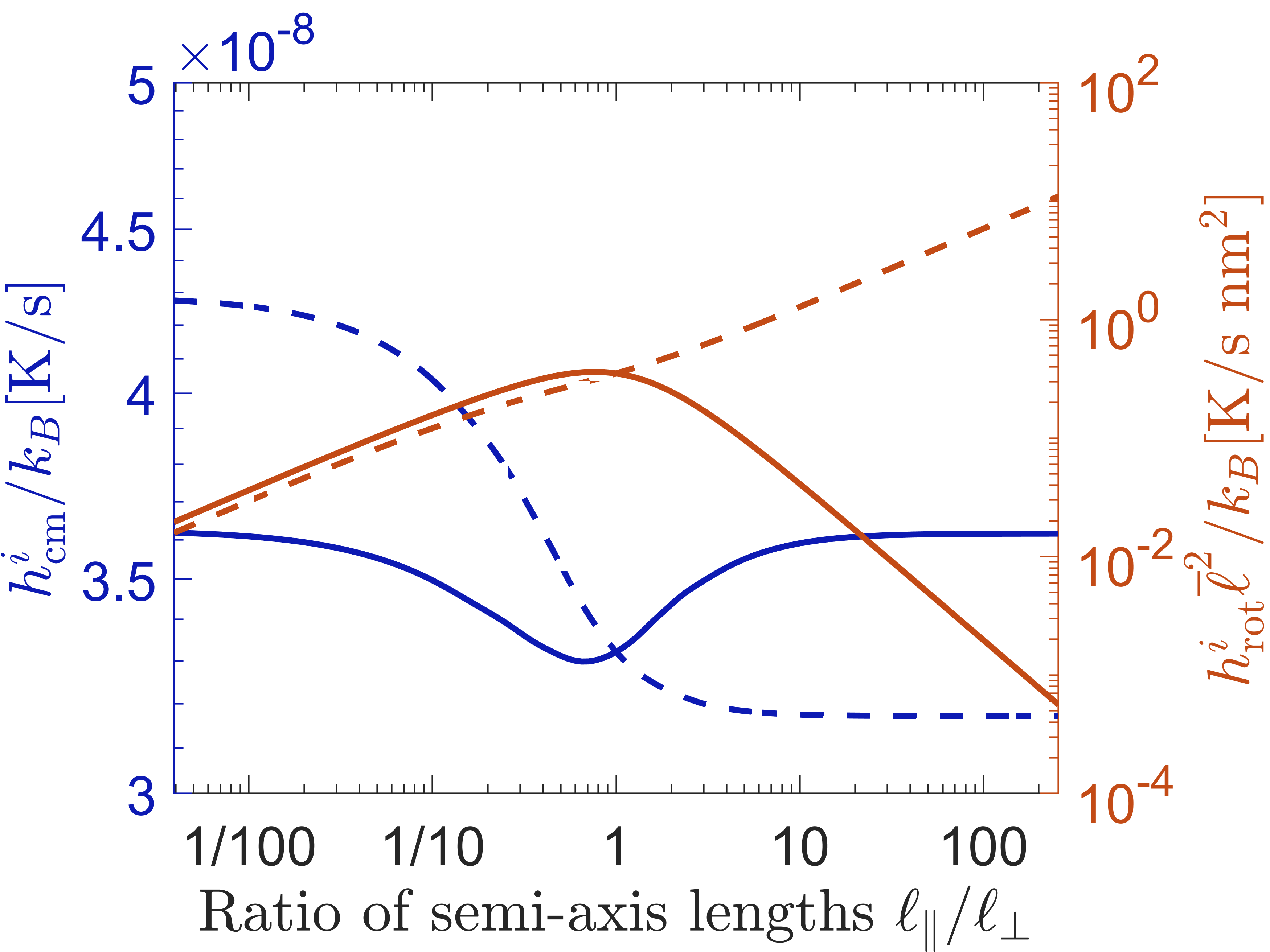}};
    \begin{scope}[x={(image.south east)},y={(image.north west)}]
        \node[] at (0.04,0.98) {\textbf{(b)}};
        \node[] at (0.20,0.74) {\footnotesize$\parallel$};
        \node[] at (0.30,0.38) {\footnotesize$\perp$};
        \node[] at (0.76,0.70) {\footnotesize$\parallel$};
        \node[] at (0.72,0.34) {\footnotesize$\perp$};
    \end{scope}
\end{tikzpicture}
        }
    \caption{Motional heating rates of spheroidal submicron  silica particles  due to  thermal emission.
    The center-of-mass heating rate   $h_{\rm cm}^{i} = \partial_t \braket{P_i^2}/2 m $ (left axes, blue) is independent of size, since the small particle limit applies for all temperatures of interest. The same holds for the  librational heating rate $h_{\rm rot}^{i} = \partial_t \braket{L_i^2}/2 I_i$ when multiplied by $\bar{\ell}^{2}$ (right axes, orange), where $\bar{\ell}=(3 V /4 \pi)^{1/3}$. 
    (a) The rates, shown here for spherical particles, do not increase following a power law,  a consequence of the structured response function of realistic materials. For comparison, the dotted line shows the rate used in \cite{Chang2009}, scaling as $T^6$. 
    (b) For particles compressed or elongated along an axis the heating rates (here shown for $T=1000$\,K) differ for motion parallel (dashed) and perpendicular (solid) to the symmetry axis. The dependence on the asymmetry $\ell_\parallel/\ell_\perp$ is much weaker for the center of mass than for the rotational motion; the latter can still be estimated from (a) by multiplying the rotational heating curve of a sphere with $(\ell_\parallel/\ell_\perp)^{2/3}$ (for parallel motion) and $2 [(\ell_\perp/\ell_\parallel)^{2/3}+(\ell_\parallel/\ell_\perp)^{4/3}]^{-1}$ (for perpendicular motion).
    The calculation is based on the measured refractive index of bulk silica for wavelengths between $6\times10^{-10}$\,m and $5\times10^{-4}$\,m  at room temperature \cite{Philipp1997,Khashan2001}.
 \label{fig:Heating}
 }
\end{figure*}
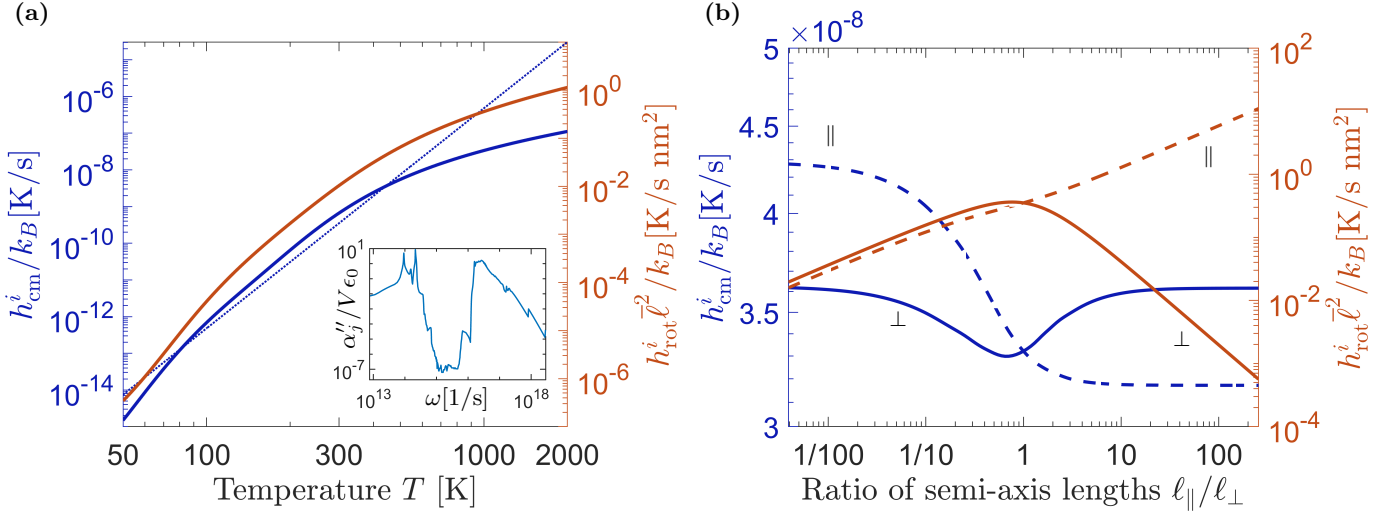

Figure~(\ref{fig:Heating}) presents the heating rates for the rotation as well as the center-of-mass motion (see Appendix \ref{app:C}) of a submicron silica spheroid. 
For example, for parameters as in the recent experiment \cite{Kamba2023} and an internal temperature of $1000\,$K this corresponds to a librational heating rate on the order of tens of phonons per second, rendering thermal emission relevant for future quantum experiments.

\subsection{General master equation}

Let us now return to the impact of heat radiation on the motional quantum state of arbitrarily sized and shaped dielectric rigid rotors.
Formulating the master equation requires  the classical electromagnetic Green tensor $\mathsf{G} (\mathbf{r}, \mathbf{r}'; \omega)$ \cite{Tai1994,Novotny2012, Buhmann2012}, which yields the electromagnetic field at position $\mathbf{r}$ sourced by a current distribution at $\mathbf{r}'$ oscillating with frequency $\omega$. 
It depends on the complex-valued relative permittivity  $\epsilon_\mathrm{r} (\mathbf{r},\omega)$, which also provides all information on the particle geometry.
Both are defined for the particle in its reference position and orientation.

For a given  source position $\mathbf{s}$,  particle orientation $\mathsf{R}$, and emission direction $\mathbf{n}$ with associated polarization vectors $\mathbf{e_\sigma}(\mathbf{n})$, we define the vector-valued amplitude of emission
\begin{align}\label{eq:KVec}
&\mathbf{K}_{{\mathsf{R}}}^{\sigma} (\mathbf{n}, \mathbf{s}; \omega) ={}
		\dfrac{1}{4 \pi} 
		\mathbf{e}_\sigma^*(\mathbf{n}) \cdot {\mathsf{R}}
		\Big[
		e^{- i \frac{\omega}{c} \mathbf{n} \cdot {\mathsf{R}} \mathbf{s}}
        \,\mathds{1}
		\\&
		+ \frac{\omega^2}{c^2} \int\limits_V \Diff{3} u \, e^{- i \frac{\omega}{c} \mathbf{n} \cdot {\mathsf{R}} \mathbf{u}} [\epsilon_\mathrm{r}(\mathbf{u}, \omega) - 1] \mathsf{G} (\mathbf{u}, \mathbf{s};\omega)
		\Big]\nonumber
		,
\end{align}
which involves the product $\mathbf{a} \cdot \mathsf{A} = \mathsf{A}^T \mathbf{a}$ and an integral over the particle volume $V$ (in reference position and orientation).  The vectors $\mathbf{n}$, $\mathbf{e}_1$, $\mathbf{e}_2$ form an orthonormal set (with respect to the complex scalar product $\mathbf{a}^*\cdot\mathbf{b}$).
(Note that the integration over the singularity of the Green tensor has to be carried out with care \cite{Novotny2012}.)

The master equation then takes the form 
\begin{align}\label{eq:MEGen}
		\mathcal{D} \rho
		={}& \!
		\int\limits_{0}^{\infty} \diff \omega \int\limits_{S^2} \Diff{2} n \int\limits_{V} \Diff{3} s \! \! \sum\limits_{\sigma \in \{1,2\} }
		\dfrac{2 \omega^3}{\pi c^3}
		\overline{n} ( \mathbf{s}, \omega)
        \text{Im} [\epsilon_\mathrm{r} (\mathbf{s}, \omega ) ]
		\nonumber\\&
  \times
		\bigg[
		e^{- i \frac{\omega}{c} \mathbf{n} \cdot \hat{\mathbf{X}}} \, \mathbf{K}_{\hat{\mathsf{R}}}^{\sigma}(\mathbf{n}, \mathbf{s}; \omega)
  \cdot
		\, \rho \,
		\mathbf{K}_{\hat{\mathsf{R}}}^{\sigma*}
  (\mathbf{n}, \mathbf{s}; \omega) \,
  e^{i \frac{\omega}{c} \mathbf{n} \cdot \hat{\mathbf{X}}} 
		\nonumber\\
		&
		- | \mathbf{K}_\mathds{1}^{\sigma}(\mathbf{n}, \mathbf{s}; \omega) |^2 \rho
		\bigg]
  .
\end{align}
To see that Eq.~(\ref{eq:MEGen}) is of Lindblad form,  consider the vector jump operators $\hat{\mathbf{L}}_\sigma (\mathbf{n}, \mathbf{s}, \omega)
=
 e^{- i \frac{\omega}{c} \mathbf{n} \cdot \hat{\mathbf{X}}} \,
\mathbf{K}_{\hat{\mathsf{R}}}^{\sigma} (\mathbf{n}, \mathbf{s}; \omega)$, which inherit their operator-valuedness from the position vector $\hat{\mathbf{X}}$ and rotation tensor $\hat{\mathsf{R}}$, and note that $\text{Im} \, \epsilon_\mathrm{r} (\omega, \mathbf{s}) > 0$.
The anti-commutator appearing in the Lindblad equation then reduces to a number when integrated over all emission directions, since $ \int_{S^2} \Diff 2 n |\hat{\mathbf{L}}_\sigma (\mathbf{n}, \mathbf{s}, \omega ) |^2=  \int_{S^2} \Diff 2 n | \mathbf{K}_\mathds{1}^{\sigma}(\mathbf{n}, \mathbf{s}; \omega) |^2$, giving the last line in (\ref{eq:MEGen}).

The operators $\hat{\mathbf{L}}_\sigma (\mathbf{n}, \mathbf{s}, \omega)$ can be interpreted as amplitudes describing how a polarization fluctuation at position $\mathbf{s}$ contributes  to the emission of an asymptotically free photon with wavevector $\omega \mathbf{n}/c$ and polarization $\mathbf{e}_\sigma$. Their dependence on the position and orientation observables thus accounts for the entanglement between the emitted radiation and the motional degrees of freedom.
The corresponding rate is determined by the imaginary part of the dielectric function and the Bose-Einstein occupation factor, in accordance with the fluctuation-dissipation theorem  (see Sect.~\ref{chap:PhEm}). This reflects the fact that emission is driven by the thermal fluctuations of the polarization currents in the material.
The particle temperature may vary locally as long as the assumption of quasi-equilibrium holds.

Notice that the center-of-mass position
appears  only in terms of momentum kick operators in Eq.~(\ref{eq:MEGen}), which describe the linear momentum recoil due to photon emission. In contrast, the orientation observable 
appears in Eq.~(\ref{eq:KVec}) both in the exponents and in front of the square brackets, as required when rotating the emitted radiation vector field to the reference orientation.
The exponential functions can be interpreted as imparting (superpositions of) angular momentum kicks according to the lever arm pointing from the center of mass to the last point of interaction within the particle, compensating the orbital angular momentum carried away by the emitted photon.

The master equation (\ref{eq:MEGen}) acts as multiplication by a complex rate when expressed in position-orientation basis, see  Eq.~(\ref{eq:Drho}), because all jump operators are  functions of only the position and orientation observables.
By completing the square as $2 a b^* - |a|^2 - |b|^2 = -|a-b|^2 + 2 i \, \text{Im}[ a b^*]$ for each component of $\mathbf{K}_\mathsf{R}^{\sigma} (\mathbf{n}, \mathbf{s}; \omega)$, the rate takes the form
\begin{align}\label{eq:Ftilde}
		& {F}_{\mathsf{R},\mathsf{R}'} (\Delta \mathbf{X})
		={}
  \nonumber\\&
		\int\limits_{0}^{\infty} \diff \omega \int\limits_{S^2} \Diff{2} n \int\limits_{V} \Diff{3} s \! \! \sum\limits_{\sigma \in \{1,2\} }
		\dfrac{2 \omega^3}{\pi c^3}
        \overline{n} (\mathbf{s},  \omega )
		\text{Im} [\epsilon_\mathrm{r} (\mathbf{s}, \omega) ]
		\nonumber\\&
  \times
		\bigg(
            \dfrac{1}{2} \Big|
            e^{- i \frac{\omega}{c} \mathbf{n} \cdot \Delta \mathbf{X}} \, 
            \mathbf{K}_{\mathsf{R}}^{\sigma}
            (\mathbf{n}, \mathbf{s}; \omega)
            -
            \mathbf{K}_{\mathsf{R}'}^{\sigma}
            (\mathbf{n}, \mathbf{s}; \omega)
            \Big|^2
            \nonumber\\&
            + i \mathrm{Im} \Big[
            e^{- i \frac{\omega}{c} \mathbf{n} \cdot \Delta \mathbf{X}}
            \mathbf{K}_{\mathsf{R}}^{\sigma}
            (\mathbf{n}, \mathbf{s}; \omega)
            \cdot
            \mathbf{K}_{\mathsf{R}'}^{\sigma*}
            (\mathbf{n}, \mathbf{s}; \omega)
            \Big]
		\bigg)
  .
\end{align}

The real part, which is non-negative, is called localization rate because it describes an exponential decay of the positional and orientational coherences.
Note that it will generally be finite even if the flux of outgoing photons into a given direction (which is proportional to $|\mathbf{K}_\mathsf{R}^{\sigma} (\mathbf{n}, \mathbf{s}; \omega)|^2$) is independent of the particle position and orientation. 
This can be understood as a result of phase information being contained in the emitted radiation. 

Equation (\ref{eq:Ftilde}) shows that the localization rate is bounded by twice the emission rate (\ref{eq:PhFluxGen}), since $ ||\mathbf{a} - \mathbf{b}||^2/2 \le ||\mathbf{a}||^2 + ||\mathbf{ b} ||^2 $. In the limit of large delocalizations, $\Delta X \to \infty$,  it converges to the emission rate because the  oscillations of the mixed terms average out.
Moreover, the definition (\ref{eq:KVec}) implies 
${F}_{\mathsf{R},\mathsf{R}'} (\Delta \mathbf{X}) =  {F}_{\mathds{1}, \mathsf{R}^{ T} \mathsf{R}'} (\mathsf{R}^{ T} \Delta \mathbf{X})$. If both positional and orientational superpositions are present, the loss of coherence is therefore not just a function of the relative position $\Delta \mathbf{X}$ and relative orientation $\mathsf{R}^T \mathsf{R}'$.

\subsection{Small and large particle limits}

The decoherence rate (\ref{eq:Ftilde})
simplifies if the particle  size is much smaller or much greater than both the thermal wave length and the corresponding attenuation length in the bulk material.

In the limit of small particles, multiple photon scattering in the particle can be neglected, while the electrostatic interaction of the thermally fluctuating dipoles must still be accounted for.
Assuming a homogeneous permittivity and the particle shape to be ellipsoidal, an integral equation for $\mathbf{K}_{{\mathsf{R}}}^{\sigma} (\mathbf{n}, \mathbf{s}; \omega)$ can 
then be solved self-consistently, as described in App.~\ref{chap:SPlim}. The solution depends on a limit of the free-space Green tensor $\mathsf{G}_0$ and on the polarizability tensor $\upalpha$, see Eq.~(\ref{eq:KPP}). Insertion into (\ref{eq:MEGen}) yields the master equation (\ref{eq:MEPPFull}).
The corresponding localization rate can be simplified to
\begin{align}\label{eq:SPDecRaFull}
		F_{\mathsf{R},\mathsf{R}'} ( \Delta \mathbf{X} )
		={}&
		\int\limits_{0}^{\infty} \diff \omega \,
  \dfrac{\omega^3}{\pi^2 c^3 \epsilon_0 } \overline{n} (\omega) \,
		\TR \bigg[
  \dfrac{\upalpha'' (\omega)}{3}
		\nonumber\\&\times
		\left( \mathds{1} - \dfrac{6 \pi c}{\omega}\mathsf{R}'^{T} \text{Im} \, \mathsf{G}_0 (\Delta \mathbf{X}; \omega)
		\mathsf{R} \right) \bigg], 
\end{align}
which  depends only on  the imaginary parts of  the polarizability and the  free-space Green tensor.
The positivity of (\ref{eq:SPDecRaFull}) follows from the Lindblad form of  (\ref{eq:MEPPFull}) (and can be shown alternatively  using the von Neumann trace inequality and the Hölder inequality, as well as bounds for the eigenvalues of $\text{Im} \, \mathsf{G}_0 (\Delta \mathbf{X}; \omega)$).
The special case of a linear rotor is given in App.~\ref{app:A}.

Figure \ref{fig:Main}(b) shows  the  spatio-orientational decoherence rate for superpositions involving center-of-mass translations by $\Delta X $ (in a direction $\mathbf{m}$) and rotations by $\theta$ (around the same $\mathbf{m}$). We consider the  case of a sphere, where the localization rate is independent  of $\mathbf{m}$.
Note that for general superpositions the decoherence rate cannot be decomposed into a function of the reduced states of center of mass and orientation. 

In the limit of large particles, the decoherence rate  (\ref{eq:Drho}) scales with the surface of the particle, as one would expect from the scaling of heat radiation. Specifically, assuming that the particle extension and radii of curvature are much greater than wavelength and absorption length, and its surface to be convex, the spectral photon intensity emitted from each surface element (spectral rate of photons per solid angle per surface area) reads \cite{Eckhardt1984}
\begin{align}\label{eq:Phis}
\Phi_\mathrm{s} (\mathbf{n}, \mathbf{r}_\mathrm{s}; \omega)
={}
		\dfrac{\omega^2}{8 \pi^3 c^2 } \overline{n} (\omega)
		[T_\mathrm{s} (\mathbf{n}, \omega) + T_\mathrm{p} (\mathbf{n}, \omega)]
{n^{\perp}}\Theta(n^{\perp}) 		.
\end{align}
Here, we dropped the dependencies on 
the position vector $\mathbf{r}_\mathrm{s}$ of the surface element for brevity. $T_\mathrm{s}$ and  $T_\mathrm{p}$ are the Fresnel transmission coefficients for radiation \textit{incident} on the particle, and $n^\perp(\mathbf{r}_\mathrm{s})$ is the component of the emission direction $\mathbf{n}$ normal to the local surface
(the Heaviside function ensures that radiation is only emitted away from the body).
The dependence on the transmission of incident radiation can be considered  a manifestation of Kirchhoff’s law of thermal
radiation,  while the explicit $n^{\perp}$-dependence yields Lambert’s emission law in the case of a black body, where $T_\mathrm{s}=T_\mathrm{p}=1$.

For large particles, the volume integral in (\ref{eq:MEGen}) reduces to an integration over a surface shell with an effective depth given by the absorption length (see App.~\ref{chap:LPlim}). The associated  master equation (\ref{eq:bigone}) simplifies considerably for well oriented states such that  $\langle\mathsf{R}|\rho|\mathsf{R}'\rangle \neq 0$
only for $||\mathsf{R}^{(\prime)}-\mathds{1}||\ll 1$. In this case, the orientational decoherence is  dominated by orbital angular momentum kicks, and the master equation (\ref{eq:MEGen})   takes the form
\begin{align}\label{eq:LP}
		\mathcal{D} \rho ={}&
		\int\limits_{0}^{\infty} \diff \omega \int\limits_{S^2} \Diff 2 n \int\limits_{\partial V} \Diff 2 r_\mathrm{s} \,
        \Phi_\mathrm{s} (
        \mathbf{n}, \mathbf{r}_\mathrm{s}, \omega)
  \nonumber\\&\times
		\Big[
		e^{- i \frac{\omega}{c} \mathbf{n} \cdot ( \mathbf{X} + \mathsf{R} \mathbf{r}_\mathrm{s})}
		\rho
		e^{i \frac{\omega}{c} \mathbf{n} \cdot ( \mathbf{X} + \mathsf{R} \mathbf{r}_\mathrm{s})}
		- \rho
		\Big]
		.
	\end{align}
The jump operators are now a tensor product of two unitaries, so that the reduced dynamical equations of the center-of-mass and  the orientation state are of closed form,  readily obtained by a partial trace over (\ref{eq:LP}).

\section{Derivation}

Standard methods are unsuitable for obtaining the motional decoherence of extended bodies due to thermal radiation.
This is because emission will occur only at frequencies where the  dielectric function has a finite imaginary part, a consequence of the fluctuation-dissipation theorem for the thermally driven polarization currents.
The thus unavoidable presence of absorption in the medium implies that the macroscopic electromagnetic field cannot be decomposed into modes, 
so that the usual approaches for deriving the open quantum dynamics of motional degrees of freedom fail.
Instead, the master equation (\ref{eq:MEGen}) is obtained in two steps.

First, we derive the Lindblad equation describing how the ro-translational state of an extended body is affected by the probabilistic emission of scalar particles.
The  jump operators quantifying the associated linear and angular momentum kicks can be expressed in terms of the Helmholtz Green function. 
Second, we move on to the probabilistic emission of photons by replacing the Green function of the scalar equation with the one for the vector Helmholtz equation. The emission rates are then obtained from the theory of fluctuation electrodynamics.

\subsection{Emission of scalar particles}

We start by considering the emission of weakly bound scalar particles of mass $m$, referred to as `atoms' in the following. Their decay from a metastable state  can be modeled by introducing an internal two-level system. For $\ket{\uparrow}$ the atom is in the bound state $|\varphi_0\rangle$, while it evolves freely in the potential $V(\mathbf{r})$ for $\ket{\downarrow}$. The emission dynamics of the initially bound state  $\ket{\Psi (0)} = \ket{\varphi_0} \ket{\uparrow} $  is then described by the Hamiltonian
\begin{align}
		H_\mathrm{tot} ={}&
  E_0 \proj{\varphi_0}\otimes \proj{\uparrow} + H\otimes \proj{\downarrow}
  \nonumber\\&
  + g \hbar
  \proj{\varphi_0}\otimes (\ket{\uparrow}\bra{\downarrow}+\ket{\downarrow}\bra{\uparrow})
\end{align}
where $H = p^2/2m + V$
and $g$ is the coupling rate. 
Expanding the state  as
\begin{equation}
		\ket{\Psi(t)} ={}
		e^{- i E_0 t /\hbar} b(t) \ket{\varphi_0} \ket{\uparrow} + \int\limits_{0}^{\infty} \diff E \, c_E (t) e^{- i E t /\hbar} \ket{E}\ket{\downarrow}
\end{equation}
and assuming the coupling to be small, we obtain the dynamics of the bound part $b(t)$  in a Born-Markov-like approximation  by inserting the formal solution of $ c_E (t)$ into the dynamical equation for $b(t)$, replacing $b(t') $ by $ b(t)$, extending the integration limit to infinity, and neglecting the Lamb-shift-like  renormalization of $E_0$.
This yields an exponential decay $ b(t)= e^{- \Gamma_0 t /2} $ of the bound part with rate $ \Gamma_0 = 2 \pi \hbar g^2 |\braket{E_0 | \varphi_0}|^2 $. Inserting this into the equation for $c_E(t)$ yields the wave function $\ket{\psi_\downarrow}=\braket{\downarrow | \Psi(t)}$ of the unbound part  for $\Gamma_0 t\ll 1$
 \begin{align}
		\ket{ \psi_\downarrow(t)} ={}
		- \hbar g \int\limits_{0}^{\infty} \diff E 
		\big(
		e^{- i E_0 t /\hbar }
		  + e^{- i E t /\hbar}
		\big)
		\dfrac{\ket{E}\braket{E|\varphi_0}}{E_0 - i \hbar \Gamma_0 /2 - E}
  .
\end{align}
For large $t$ the projector $\proj{E}$ can be approximated by $\proj{E_0}$ in the second term,  and the lower limit of integration can be replaced by negative infinity.
Using the Green operator $G(z) = [z - H]^{-1}$ and the relation $\ket{E}\bra{E}=i[G(E+i\epsilon)-G(E-i\epsilon)]/2\pi$  then leads to
	\begin{align}\label{eq:psidown}
\ket{\psi_\downarrow}=  \lim\limits_{\Gamma_0 \to 0} \hbar g e^{- i E_0 t /\hbar} G (E_0 + i \hbar \Gamma_0/2) \ket{\varphi_0}
  .
\end{align}
The emitted wavefunction $\ket{\psi_\downarrow}$ can therefore be understood as the retarded solution to the inhomogeneous Schrödinger equation $(i\hbar\partial_t - {H})\ket{\psi_\downarrow}=\hbar g \exp(- i E_0 t /\hbar)\ket{\varphi_0}$. 

It will be useful below to define amplitudes of emission
\begin{equation}\label{eq:DefScalarA}
    A (\mathbf{n},\mathbf{s}, \omega) =
    -
    \lim_{r \to \infty} r e^{- i r p/\hbar} G (r \mathbf{n}, \mathbf{s}; \omega)
\end{equation}
in terms of  the retarded Green function $G(\mathbf{r},\mathbf{r}'; E/\hbar)=\bra{\mathbf{r}}G(E+i\epsilon)\ket{\mathbf{r}'}$, where $p=\sqrt{2m\hbar\omega}$.
These amplitudes describe the asymptotic form of the wavefunction in direction $\mathbf{n}$ due to a source at  position $\mathbf{s}$. They can be expressed as matrix elements of the M\o{}ller operators $\Omega_{\pm} = \lim_{t \to \mp \infty} e^{ i H t / \hbar} e^{ - i H_0 t / \hbar}$ associated with scattering off the potential $V=H-H_0$,
\begin{align}\label{eq:Anew}
		A ( \mathbf{n}, \mathbf{s}, \omega)={}
    \dfrac{1}{4 \pi}
    \dfrac{2m}{\hbar^2}
    (2 \pi \hbar)^{3/2}
  \braket{\mathbf{n} p |
  \Omega_-^\dagger |
    \mathbf{s}}
    .
\end{align}
Starting from the right hand side, we used that eigenstates of $H_0$ are mapped to $\Omega_{\pm} \ket{E}_0 =		\lim_{\epsilon \to 0} \pm i \epsilon G(E \pm i \epsilon) \ket{E}_0$, inserted a resolution of identity in position space, used that finite domains of integration do not contribute, replaced $G$ by its asymptotic form, and performed a stationary phase approximation on the solid angle integration. 
Moreover, we dropped the energy dependence for brevity.

\subsection{Inclusion of the motional degrees of freedom}

We next take into account that
the emission process
depends on the particle position $\mathbf{X}$ and orientation $\mathsf{R}$. For a fixed spatio-orientational state $\ket{\mathbf{X},\mathsf{R}} $ this is accomplished by the unitary operator $D (\mathbf{X}, \mathsf{R})$, which transforms the emission problem from the reference position and orientation of the particle to the actual one.
Conditioned on an emission having taken place, the transformation of the motional state $\rho$
is given by
\begin{equation}
    \rho' = \tr_{\mathrm{atom}} \big[ D (\hat{\mathbf{X}}, \hat{\mathsf{R}}) \proj{\psi_\downarrow} \otimes \rho D^\dagger (\hat{\mathbf{X}}, \hat{\mathsf{R}})
    \big]
    ,
\end{equation}
where
$ D (\hat{\mathbf{X}}, \hat{\mathsf{R}}) \ket{\mathbf{r}} \ket{\mathbf{X}, \mathsf{R} } = \ket{\mathsf{R} \mathbf{r} + \mathbf{X}} \ket{\mathbf{X}, \mathsf{R}} $.

Inserting (\ref{eq:psidown}) with $g^2=\Gamma_0 /( 2 \pi \hbar|\braket{E_0 | \varphi_0}|^2 ) $, 
using the resolvent equation $G=(\mathds{1}+G V)G_0$ with  $G_0 (z) = (z-H_0)^{-1}$, 
carrying out the trace  in momentum basis, 
identifying the nascent delta function $\delta(E_0-p^2/2m)$ as $\Gamma_0 \to 0$, which enforces the on-shell momentum $p_0 = \sqrt{2 m E_0}$, and recognizing the  M\o{}ller-out operators $\Omega_-$
one arrives at
	\begin{align}\label{eq:CondState}
		\rho' ={}&
		\dfrac{p_0 m}{|\braket{E_0|\varphi_0}|^2}
		\int\limits_{S^2} \Diff2n \,
		\braket{p_0 \mathbf{n} |
		\hat{D}
		\Omega_{-}^\dagger
		|\varphi_0}
		\rho
		\braket{\varphi_0 |
			\Omega_{-}
			\hat{D}^\dagger
			|p_0 \mathbf{n} } .
	\end{align}
Here we abbreviated $\hat{D} \equiv D(\hat{\mathbf {X}},\hat{\mathsf{R}})$, and we note that the emission probability vanishes for $\braket{E_0|\varphi_0} = 0$.
The normalization is preserved   
since
\begin{equation}\label{eq:OrIntToE}
	m p_0\int\limits_{S^2} \Diff2n  \,\Omega_-\proj{p_0 \mathbf{n}}\Omega_-^\dagger = \proj{E_0}.
\end{equation}
In the following we assume the bound states to be  sufficiently localized  so that the $\ket{\varphi_0}$ in Eq.~(\ref{eq:CondState}) may be replaced by a position eigenstate. 
Accordingly, to keep $\Gamma_0$ fixed the coupling rate
is replaced, $g^2 \to \tilde{g}^2 = \Gamma_0 / (2 \pi \hbar |\braket{E_0| \mathbf{r}_0}|^2 )$.

The asymptotic probability flux of the emitted atom per solid angle in direction $\mathbf{n}$ depends on the orientation observable $\hat{\mathsf{R}}$ of the particle. It is given by
\begin{align}\label{eq:MassiveFlux}
  \hat{\Phi} (\mathbf{n})
  ={}&
		\lim\limits_{r\to \infty} r^2 \mathbf{n} \cdot \dfrac{1}{m}  \text{Re} \left[
		\braket{\psi_\downarrow | \hat{D}^\dagger \delta (r \mathbf{n} - \mathbf{x}) \mathbf{p} \hat{D} | \psi_{\downarrow}} 
		\right]
		\nonumber\\={}&
		\Gamma_0
		\dfrac{
			\left|
   \braket{p_0  \mathbf{n} | \hat{D} \Omega_-^\dagger | \mathbf{r}_0}
			\right|^2}
		{\int_{S^2}\Diff2n' \,
			| \braket{p_0 \mathbf{n}' | \Omega_-^\dagger | \mathbf{r}_0}|^2}
   \nonumber\\={}&\hbar^2\tilde{g}^2
		\dfrac{ p_0 }{m} 
		\big|
		A ( \hat{\mathsf{R}}^T\mathbf{n}, \mathbf{r}_0)
		\big|^2
  .
\end{align}
The second line confirms that $\Gamma_0$ is the total atom emission rate. 

\subsection{Construction of the master equation}

The master equation governing the reduced spatio-orientational dynamics on a coarse-grained timescale $\diff t$  can now be obtained by applying the transformation Eq.~(\ref{eq:CondState}) with probability $ \Gamma_0 \diff t$, while taking into account the possibility of non-emission with probability $ 1- \Gamma_0 \diff t$.
Moreover, we allow for a distribution $w(\mathbf{s},\omega)$ of bound atoms with varying 
locations $\mathbf{s}$, bound energies $\hbar \omega$, and couplings $\tilde{g}(\mathbf{s}, \omega)$. 
The generator of the open quantum dynamics then takes the form
\begin{align}\label{eq:allglindbladian}
    \mathcal{D} \rho ={}&
    \int\limits_0^\infty \diff \omega
    \int\limits_{S^2} \Diff2n
    \int\limits_{V} \Diff3s \,
    w(\mathbf{s},\omega)
    \hbar^2 \tilde{g}^2 (\mathbf{s}, \omega)
    \dfrac{p}{m}
    \nonumber\\&\times
    \Big[
    e^{- i p \mathbf{n} \cdot \hat{\mathbf{X}}/\hbar}
    A (\hat{\mathsf{R}}^T \mathbf{n}, \mathbf{s}, \omega)
    \rho
    A^* (\hat{\mathsf{R}}^T \mathbf{n}, \mathbf{s}, \omega)
    e^{ i p \mathbf{n} \cdot \hat{\mathbf{X}}/\hbar}
    \nonumber\\&
    \phantom{RR}
    -
    |A (\mathbf{n}, \mathbf{s}, \omega)|^2
    \rho
    \Big]
    ,
\end{align}
where $p=\sqrt{2m\hbar\omega}$.
The equation is of Lindblad form since the orientation integral over the modulus squared of the jump operators
\begin{align}\label{eq:LviaA}
\hat{ L } ( \mathbf{n}, \mathbf{s}, \omega) ={} &
e^{- i p \mathbf{n} \cdot \hat{\mathbf{X}} / \hbar }
A(\hat{\mathsf{R}}^T \mathbf{n}, \mathbf{s},\omega)
\end{align}
is a number, as follows from expansion in position-orientation basis.

If the particle is transparent to the atoms, $V(\mathbf{r})=0$, the amplitudes of emission can be given as $A_0 (\hat{\mathsf{R}}^T \mathbf{n}, \mathbf{s}, \omega) = m e^{- i p \mathbf{n} \cdot \hat{\mathsf{R}} \mathbf{s} / \hbar } / 2 \pi \hbar^2$. This operator imparts an angular momentum kick balancing out the orbital angular momentum of the outgoing particle since
$e^{- i p \mathbf{n} \cdot \hat{\mathsf{R}} \mathbf{s} / \hbar} \hat{\mathbf{J}} e^{i p \mathbf{n} \cdot \hat{\mathsf{R}} \mathbf{s} / \hbar} = \hat{\mathbf{J}} - (\hat{\mathsf{R}} \mathbf{s}) \times p \mathbf{n}$, as follows from the canonical commutation relation
$[\hat{\mathsf{R}}, \mathbf{m} \cdot \hat{\mathbf{J}} ] = i \hbar \mathbf{m} \times \hat{\mathsf{R}}$.

In the general case, one requires the  
Green function $G$ associated with the stationary Schrödinger equation for the unbounded atom in the particle potential. The  jump operators then take the form
\begin{align}\label{eq:ScalarTransformedL}
\hat{ L } ( \mathbf{n}, \mathbf{s}, \omega) ={}
- \lim\limits_{r \to \infty} r e^{- i p r /\hbar}
\,
G( \hat{\mathsf{R}}^T [r \mathbf{n} - \hat{\mathbf{X}}], \mathbf{s};\omega).
\end{align}

Moreover, the orientation-resolved spectral emission rate (\ref{eq:MassiveFlux}) per volume element can be expressed as the product of the modulus squared of the jump operators (\ref{eq:LviaA}) with their associated jump rate,
\begin{equation}\label{eq:MassiveFluxOverLs}
\hat{\Phi} (\mathbf{n}, \mathbf{s}, \omega) ={}
w (\mathbf{s}, \omega) \hbar^2 \tilde{g}^2 (\mathbf{s}, \omega) \dfrac{p}{m}
\big\vert \hat{L} (\mathbf{n}, \mathbf{s}, \omega)  \big\vert^2
.
\end{equation}

\subsection{Emission of photons}
\label{chap:PhEm}

To account for the emission of photons instead of atoms, we replace the velocity $p/m$ by the speed of light and the scalar Green function 
by the electromagnetic Green tensor. The latter is defined as the retarded solution to 
 \begin{align}\label{eq:Gtensor}
		\left[ \nabla \times
  \nabla \times - \dfrac{\omega^2}{c^2} \epsilon_\mathrm{r} (\mathbf{r}, \omega) \right]
		\mathsf{G} (\mathbf{r}, \mathbf{r'}; \omega ) ={} \mathds{1} \delta (\mathbf{r} - \mathbf{r}').
\end{align}
It determines the electromagnetic field at position $\mathbf{r}$ due to a (polarization) current density $\mathbf{j}(\mathbf{r}', \omega)$ 
oscillating with frequency 
$\omega$
\cite{Tai1994, Novotny2012, Buhmann2012}
\begin{align}
    \mathbf{E} (\mathbf{r}, \omega) ={}& i \mu_0 \omega
    \int\limits_{\mathds{R}^3} \Diff 3 r'\,
    \mathsf{G} (\mathbf{r}, \mathbf{r}'; \omega) \,
    \mathbf{j} (\mathbf{r}', \omega)
    \\
    \mathbf{B} (\mathbf{r}, \omega) ={}& \mu_0
    \int\limits_{\mathds{R}^3} \Diff 3 r'\,
    \nabla \times
    \mathsf{G} (\mathbf{r}, \mathbf{r}'; \omega) \,
    \mathbf{j} (\mathbf{r}', \omega)
    .
\end{align}
Here, we assume the particle  to be composed of a non-magnetic dielectric material described in reference position and orientation by the relative permittivity $\epsilon_\mathrm{r} (\mathbf{r}, \omega)$.

The amplitudes of emission (\ref{eq:DefScalarA}) are now tensorial,
\begin{equation}\label{eq:DefVecA}
  \mathsf{A}
  (\mathbf{n}, \mathbf{s}, \omega) ={}
		\lim\limits_{r \to \infty} r e^{- i \frac{\omega}{c} r}
  \,
  \mathsf{G} (r \mathbf{n}, \mathbf{s} ; \omega)
  ,
\end{equation}
so that 
the jump operators (\ref{eq:LviaA}) take the form
\begin{align}\label{eq:DefVecL}
\hat{ \mathsf{L} } ( \mathbf{n}, \mathbf{s}, \omega) ={}&
e^{- i \frac{\omega}{c} \mathbf{n} \cdot \hat{\mathbf{X}} } \, 
\hat{\mathsf{R}} \mathsf{A}(\hat{\mathsf{R}}^T \mathbf{n}, \mathbf{s}, \omega)
,
\end{align}
where the additional rotation tensor $\hat{\mathsf{R}} $  accounts for the vectorial nature of the emitted radiation.
Note that the definition  (\ref{eq:DefVecA}) complies with the different sign convention for $\mathsf{G}$ as compared to $G$ in (\ref{eq:DefScalarA}).

We obtain the flux of thermally emitted photons by first calculating the asymptotic Poynting vector using the theory of fluctuation electrodynamics, where the field statistics are governed by the fluctuation-dissipation theorem \cite{Rytov, Krueger2012, Buhmann2012}.
For pure dielectrics the two-point correlator of the polarization current density can be given in compact form as \cite{Eckhardt1982, Novotny2012}
\begin{align}
		\langle {j}_{i} (\mathbf{r}, \omega) {j}_{j}^* (\mathbf{r}', \omega') \rangle ={}&
		\frac{\hbar \omega^2 \epsilon_0 }{\pi}\text{Im} [ \epsilon_\mathrm{r} (\mathbf{r}, \omega)]
            \big[\overline{n} (\mathbf{r}, \omega) + \dfrac{1}{2}\big]
            \nonumber\\&\times
		\delta_{i j} \delta (\omega - \omega') \delta(\mathbf{r} - \mathbf{r'})
.
\end{align}

Using the real-valuedness of the fields in the time domain and an integral identity of the Green tensor, here for nonmagnetic media,
\begin{align}
&
\dfrac{\omega^2}{c^2} 
\int\limits_{\mathds{R}^3} \Diff 3 s \,
\nabla \times \mathsf{G}(\mathbf{r}, \mathbf{s} ; \omega) \mathrm{Im} [\epsilon_\mathrm{r} ( \mathbf{s}, \omega)] \mathsf{G}^\dagger (\mathbf{r}', \mathbf{s}; \omega)
\nonumber\\
&=
\mathrm{Im} \big[ \nabla \times \mathsf{G} (\mathbf{r}, \mathbf{r}'; \omega) \big]
\end{align}
enables one to eliminate the vacuum fluctuations \cite{Eckhardt1984},
and with the Silver-Müller radiation condition \cite{
Schot1992, Tai1994} the asymptotics of the Poynting vector reads
\begin{align}\label{eq:PoyMQEDasy}
		\braket{
  \mathbf{S} (r \mathbf{n})
  }
		\sim{}
		\mathbf{n}
  &
		\int\limits_{0}^{\infty} \diff \omega \int\limits_{V} \Diff{3} s \,
		\dfrac{2 \omega^3}{\pi c^3} \hbar \omega \overline{n} (\mathbf{s},  \omega ) \text{Im} [ \epsilon_\mathrm{r} ( \mathbf{s}, \omega )]
  \nonumber\\&\times
  \TR \big[
		\mathsf{G} ( r \mathbf{n}, \mathbf{s}; \omega)
  \mathsf{G}^\dagger ( r \mathbf{n}, \mathbf{s}; \omega)
  \big]
		,
	\end{align}
 as $r\to\infty$.
The spectral flux of emitted photons per solid angle and volume element is thus given by
\begin{align}\label{eq:PhFluxGen}
		\Phi (\mathbf{n}, \mathbf{s}, \omega) ={}&
		\dfrac{2 \omega^3}{\pi c^3} \overline{n} (\mathbf{s},  \omega ) \text{Im} [ \epsilon_\mathrm{r} ( \mathbf{s}, \omega ) ]
  \TR \big[
		\mathsf{A} (\mathbf{n}, \mathbf{s}, \omega) \mathsf{A}^\dagger (\mathbf{n}, \mathbf{s}, \omega)
  \big]
  .
\end{align}
This holds for the particle in reference orientation; for arbitrary orientational states the photon flux operator $\hat{\Phi} (\mathbf{n}, \mathbf{s}, \omega) = \Phi (\hat{\mathsf{R}}^T \mathbf{n}, \mathbf{s}, \omega)$ can again be expressed in terms of the jump operators,
\begin{align}\label{eq:PhFluxGenL}
		\hat{\Phi} (\mathbf{n}, \mathbf{s}, \omega) ={}&
		\dfrac{2 \omega^3}{\pi c^3} \overline{n} (\mathbf{s},  \omega ) \text{Im} [ \epsilon_\mathrm{r} ( \mathbf{s}, \omega ) ]
  \TR \big[
		\hat{\mathsf{L}} (\mathbf{n}, \mathbf{s}, \omega) \hat{\mathsf{L}}^\dagger (\mathbf{n}, \mathbf{s}, \omega)
  \big]
  .
\end{align}
Comparison with Eq.~(\ref{eq:MassiveFluxOverLs}) shows that the  jump rate associated with the $\hat{\mathsf{L}} (\mathbf{n}, \mathbf{s}, \omega)$ is given by $2 \omega^3 \overline{n} (\mathbf{s},  \omega ) \text{Im} \epsilon_\mathrm{r} ( \omega, \mathbf{s} )/ \pi c^3$, which concludes the derivation of the master equation.

To arrive at the
Eqs.~(\ref{eq:KVec}), (\ref{eq:MEGen}), one may use the (right-hand) Dyson equation \cite{Buhmann2013, Martin1995Jan}
\begin{align}\label{eq:DysonED}
		\mathsf{G} (\mathbf{r}, \mathbf{r}' ; \omega)
		&={}
  \mathsf{G}_0 (\mathbf{r}- \mathbf{r}'; \omega)
  \nonumber\\&
  + \dfrac{\omega^2}{c^2} \int\limits_{\mathds{R}^3} \Diff{3} s \,
		\mathsf{G}_0 (\mathbf{r}- \mathbf{s}; \omega) [\epsilon_\mathrm{r} (\mathbf{s}, \omega) -1 ] \mathsf{G} (\mathbf{s}, \mathbf{r}'; \omega)
		,
\end{align}
before taking the limit $r \to\infty$ in the definition (\ref{eq:DefVecA}).
Here $\mathsf{G}_0  $ is the free-space Green tensor \cite{Novotny2012}
\begin{align}\label{eq:G0}
		\mathsf{G}_0 (r \mathbf{e}_r; c k) ={}&
		\dfrac{e^{i k r}}{4 \pi r}
		\bigg[ \left( 1 + \dfrac{i k r - 1}{k^2 r^2}\right) \mathds{1}
		\nonumber\\&
  + \dfrac{3 - 3 i k r - k^2 r^2}{k^2 r^2} \mathbf{e}_r \otimes \mathbf{e}_r\bigg]
.
\end{align}
To finally perform the tensorial trace one may use that $\mathbf{n} \cdot \mathsf{A} (\mathbf{n}, \mathbf{s}, \omega) = 0$, as follows from $\mathsf{G}$ being asymptotically transverse. This leaves a sum over polarizations $\sigma\ \in \{1,2\}$ involving the vector functions 
\begin{align}\label{eq:Kvecdef}
    \mathbf{K}^{\sigma}_\mathsf{R}(\mathbf{n},\mathbf{s}; \omega) = \mathbf{e}_\sigma^* \cdot \mathsf{R} \mathsf{A} ( \mathsf{R}^T \mathbf{n}, \mathbf{s}, \omega)
\end{align} which appear in Eq.~(\ref{eq:MEGen}).

\section{Discussion}

The master equation (\ref{eq:MEGen}) serves to predict the spatio-orientational decoherence of arbitrarily shaped and sized dielectric particles due to the emission of thermal radiation. It holds provided that the internal degrees of freedom remain in quasi equilibrium, described by a local temperature. Their thermal fluctuations drive a polarization current, generating electromagnetic radiation that may be scattered or reabsorbed inside the particle. The radiation leaving the body conveys information on the whereabouts and alignment of the particle, leading to decoherence of its spatio-orientational quantum state. Equivalently, the impact of the 
heat radiation can be understood 
in terms of the  linear and angular momentum carried away by each outgoing photon. However, unlike for the center of mass, the effect of a photon emission event on the orientation state is more involved than a simple angular momentum kick, imposing the complicated form of Eq.~(\ref{eq:MEGen}).

In the limiting cases of very small or very large particles the master equation reconfirms previous results and physical expectations. For particles much smaller than the thermal wavelength orbital angular momentum kicks play no role and the decoherence effect can be fully described in terms of the polarization tensor. In this case, one may recover the center-of-mass decoherence rates given in \cite{Chang2009,Bateman2014} 
by  assuming the rotation state to remain uncorrelated with the center-of-mass state. (Ref.\ \cite{Chang2009} differs by a factor of two, and \cite{Bateman2014} by replacing the Bose-Einstein occupation number with the Boltzmann factor). In the opposite limit of macroscopically large bodies, decoherence turns from a volume effect to a surface effect, as one would expect in analogy to the behavior of heat radiation \cite{Bohren1983}. For well-oriented bodies the center-of-mass decoherence rates is then  controlled by the spectral photon intensity per surface area, as determined by the surface temperature and the Fresnel coefficients  (\ref{eq:Phis}).

The central ingredient of the presented theory is the spectral photon intensity sourced from each volume element of the dielectric,
which we obtain from the theory of \emph{fluctuation electrodynamics} \cite{Rytov,Krueger2012}.
The latter treats the polarization current density as a stochastic vector field, with statistics  determined by the local temperature and by the imaginary part of the dielectric function, in accordance with the fluctuation dissipation theorem.
Equivalently, the  photon intensity can be obtained from  \emph{macroscopic quantum electrodynamics} \cite{Buhmann2012} with identical results.
Photon emission rates measured for micron-scale particles agree well with the results of 
\emph{fluctuation electrodynamics}  \cite{Morino2017, Fenollosa2019, Fenollosa2023}, even in the presence of Mie resonances, so that it seems  reasonable to apply the theory to nanoparticles as well. However, since alternative predictions for the emitted heat radiation  exist in this regime \cite{Hansen1998, Lopez2018} experiments are required 
to clarify when deviations from quasi-equilibrium need to be taken into account. As direct measurements of the emitted heat radiation become increasingly difficult for smaller particles, this may require more elaborate measurement schemes \cite{Agrenius23}. In any case, a modified photon emission intensity per particle volume can be readily incorporated into the presented theory, as long as the Markov approximation holds and fluctuations remain local.

A further ingredient of the general decoherence master equation (\ref{eq:MEGen}) is the
Green tensor $\mathsf{G}$ appearing in (\ref{eq:KVec}). It encodes the scattering and reabsorption of heat radiation within the particle, and is known analytically only for simple particle geometries. However, in practice, the numerical effort of computing the Green tensor is reduced by the fact that
only
narrow frequency bands contribute significantly to heat radiation for many relevant materials, due to the first line in Eq.~(\ref{eq:MEGen}).
In particular, it suppresses the low-frequency behavior of the permittivity, which may dominate decoherence due to external thermal fluctuations \cite{Martinetz2022}.

In conclusion, the presented theory enables assessing the viability of future quantum experiments with micron-sized objects \cite{RomeroIsart2011a, scala2013, Bateman2014, 
wan2016,
Neumeier2024, RodaLlordes2024, Stickler2018,Ma2020,
Rusconi2017Oct, Rusconi2022Aug, Marletto2017Dec, Bose2017Dec, pino2018chip, martinetz2020, Weiss2021Jul, Cosco2021Jun, Rudolph2020Jan, Rudolph2022Nov, Kaltenbaek2023Jan, Schrinski2022Feb, Gasbarri2021Jul, Steiner2024May} for a given internal temperature.
If the latter proves prohibitively large, internal cooling techniques, such as laser refrigeration \cite{Rahman2017Oct, Laplane2024Mar}, must be employed. While the resulting steady-state occupations may then be inhomogeneous or non-thermal, these cases are readily incorporated into the master equation.

\section*{Acknowledgments}
JS and KH acknowledge funding by the DFG --
515993674.
BAS is supported by the DFG -- 510794108 and by the Carl Zeiss foundation through the project QPhoton.
We also acknowledge support by the Open Access Publication Fund of the University of Duisburg-Essen.

\appendix

\section{Small-particle limit}
\label{chap:SPlim}\label{app:A}

Here we derive the small-particle limit of the master equation from the  general form (\ref{eq:MEGen}) 
assuming the particle to be optically thin,  ellipsoidal, and homogeneous, $\epsilon_\mathrm{r}(\mathbf{r}\in V,\omega)=\epsilon_\mathrm{r}(\omega)$.
By self-consistently solving for the amplitudes of  emission (\ref{eq:DefVecA}), which determine the vector functions (\ref{eq:Kvecdef}), we find that
the particle properties  enter only in terms of the imaginary part of its polarizability tensor.

We start from the left-hand Dyson equation,
where the full Green tensor appears to the left of the free one, cf.~Eq.~(\ref{eq:DysonED}).
The electric susceptibility 
$\epsilon_\mathrm{r}(\mathbf{r}, \omega)-1$  restricts the spatial integration to the particle domain so that we can take the radial limit as in Eq.~(\ref{eq:DefVecA}) giving
\begin{align}\label{eq:EADyson}
	\mathsf{A} (\mathbf{n}, \mathbf{s})
	={}& \mathsf{A}_0 (\mathbf{n}, \mathbf{s}) + \dfrac{\omega^2}{c^2} [\epsilon_\mathrm{r} - 1] \int\limits_V \Diff{3} r' \,
	\mathsf{A} (\mathbf{n}, \mathbf{r}') \mathsf{G}_0 (\mathbf{r}'- \mathbf{s})
 .
\end{align}
Here, we suppress frequency dependencies for brevity.
Since Eq.~(\ref{eq:MEGen}) involves only contributions where $\mathbf{s}\in V$, and since $V$ is taken to be small compared to the attenuation length and the wave length,  the free-space Green tensor can be approximated by its asymptotic form for $ |\mathbf{r}' - \mathbf{s}| \to 0$, i.e. by the free electrostatic Green tensor.
Next, we express the latter in terms of the scalar Laplace Green function to avoid the need for a principal-value regularization of the integral \cite{VanBladel1961, Karam1997}.
Additionally, we assume the solution to the approximated integral equation to be independent of the source point $\mathsf{A}(\mathbf{n}, \mathbf{r}') = \mathsf{A} (\mathbf{n}, \mathbf{s})$.

Solving for the emission amplitude and noting that the free-space amplitude takes the form $\mathsf{A}_0 (\mathbf{n},\mathbf{s})=[\mathds{1}-\mathbf{n}\otimes\mathbf{n}]/4 \pi$ one obtains
\begin{equation}\label{eq:EADyson0}
		\mathsf{A} (\mathbf{n}, \mathbf{s})
		={} \frac{\mathds{1}-\mathbf{n}\otimes\mathbf{n}}{4\pi}
		\Big[\mathds{1} + \dfrac{\epsilon_\mathrm{r} - 1}{4 \pi}
		\nabla_s \otimes \int\limits_V \Diff 3 r'
  \nabla'
  \dfrac{1}{|\mathbf{r}' - \mathbf{s}|}
		\Big]^{-1}
  .
\end{equation}
The term in brackets is proportional to the depolarization tensor of the particle \cite{Karam1997}, which is indeed independent of $\mathbf{s}$ for ellipsoidal particles, justifying the previous assumption.

The vector functions (\ref{eq:Kvecdef}) then read
\begin{equation}\label{eq:KPP}
	\mathbf{K}_\mathsf{R}^{\sigma} (\mathbf{n}, \mathbf{s}) \approx{}
  	\dfrac{1}{4 \pi \epsilon_0 V (\epsilon_\mathrm{r}-1)}
            \mathbf{e}_\sigma^* (\mathbf{n}) \cdot \mathsf{R}\upalpha,
\end{equation}
where we identified the polarizability tensor \begin{equation}\label{eq:PolTenMED}
		\upalpha = \epsilon_0 V \sum\limits_{i=1}^{3} \mathbf{e}_i \otimes \mathbf{e}_i \dfrac{(\epsilon_\mathrm{r} -1 )}{1 + L_i (\epsilon_\mathrm{r} -1 )}
  .
\end{equation}
Here, the $\mathbf{e}_i$ are the principal axes in reference orientation. The associated depolarization factors $L_i$ depend only on the ratio of the principal 
semi-axis lengths $\{\ell_1,\ell_2,\ell_3\}$
and fulfill $L_1+L_2+L_3 = 1$.
While the exact values are given by elliptic integrals, they can be approximated as
$L_1 \approx [1 + \ell_1/\ell_2 + \ell_1/\ell_3 ]^{-1}$
and cyclic permutations
\cite{Yaghjian1980, Stratton2015, Bohren1983, Hulst1981}.

Note that solving the integral equation (\ref{eq:EADyson}) in the Born approximation would yield only the lowest order in $\epsilon_\mathrm{r} - 1$ instead of the full polarizability tensor. 
This is because the electrostatic interaction diverges too strongly at close range to be considered weak, no matter how small the particle volume.

One obtains the expression for the photoemission rates of small particles
by inserting Eq.~(\ref{eq:KPP}) into Eq.~(\ref{eq:PhFluxGen})
and using  $\upalpha'' \equiv (\upalpha - \upalpha^\dagger)/2 i
= \upalpha \upalpha^\dagger \text{Im} \epsilon_\mathrm{r} /( V \epsilon_0 |\epsilon_\mathrm{r} - 1|^2)$, as implied by (\ref{eq:PolTenMED}).

The small-particle limit of the master equation is similarly obtained by inserting (\ref{eq:KPP}) into Eq.~(\ref{eq:MEGen}). Using the transverse completeness of the polarization basis it takes the form
\begin{align}\label{eq:MEPPFull}
    \mathcal{D} \rho
		={}&
		\int\limits_{0}^{\infty} \diff \omega \int\limits_{S^2} \Diff{2} n \,
    \dfrac{\omega^3}{8 \pi^3 c^3 \epsilon_0 }
    \overline{n} (\omega) \,
		\TR \Big[ \left( \mathds{1} - \mathbf{n} \otimes \mathbf{n} \right)
  \nonumber\\&\phantom{R R R}\times
		\big(
		\hat{\mathsf{R}} \, \upalpha'' (\omega) \, e^{- i \frac{\omega}{c} \mathbf{n} \cdot \hat{\mathbf{X}}}
		\, \rho \,
		e^{ i \frac{\omega}{c} \mathbf{n} \cdot \hat{\mathbf{X}}}     \hat{\mathsf{R}}^T
        - \upalpha'' (\omega) \rho
		\big) \Big]
		,
\end{align}
where $\overline{n} (\omega)$ is the volume average over $\overline{n} (\mathbf{s}, \omega)$.
We discuss the angular momentum representation of this equation in App.~\ref{app:B}.

To finally arrive at the decoherence rate (\ref{eq:SPDecRaFull}), note that
the
solid angle integration over $e^{i k \mathbf{n} \cdot \mathbf{r}} [\mathds{1}- \mathbf{n} \otimes \mathbf{n}]$  in (\ref{eq:MEPPFull}) yields
the imaginary part of the free space Green tensor, as can be seen for instance by using plane waves as the normal modes in Eq.~(8.114) from Ref.~\cite{Novotny2012}.
Using that $\mathsf{G}_0$ and $\upalpha$ are symmetric tensors then gives
the expression shown in the main text.

The master equation governing the dynamics of a \emph{linear rotor} can be obtained from Eq.~(\ref{eq:MEPPFull})
by projecting the jump operators onto the $k=0$ subspace. This is equivalent to  replacing $ \hat{\mathsf{R}} $ by $ \hat{\mathbf{m} } \otimes \mathbf{e}_3 $, where
$\hat{\mathbf{m} } $ is the observable associated with the body-fixed symmetry axis ${\mathbf{m} } = {\mathsf{R}} \mathbf{e}_3$.
The master equation then reads as
\begin{align}\label{eq:MEPPFullLinRot}
    \mathcal{D} \rho
		={}&
		\int\limits_{0}^{\infty} \diff \omega \int\limits_{S^2} \Diff{2} n \,
    \dfrac{\omega^3}{8 \pi^3 c^3 \epsilon_0 }
    \overline{n} (\omega) \alpha_3''(\omega)
    \nonumber\\&\times
		\Big[
  e^{- i \frac{\omega}{c} \mathbf{n} \cdot \hat{\mathbf{X}}} \,
  \hat{\mathbf{m}} \cdot \left( \mathds{1} - \mathbf{n} \otimes \mathbf{n} \right)
		\, \rho \,
  \hat{\mathbf{m}}\,
		e^{ i \frac{\omega}{c} \mathbf{n} \cdot \hat{\mathbf{X}} }
        - \dfrac{2}{3} \rho
		\big) \Big]
		.
\end{align}
The associated localization rate takes the form
\begin{align}\label{eq:SPDecRaFullLinRot}
		F_{\mathbf{m},\mathbf{m}'} ( \Delta \mathbf{X} )
		={}&
		\int\limits_{0}^{\infty} \diff \omega \,
  \dfrac{\omega^3}{3 \pi^2 c^3 \epsilon_0 } \overline{n} (\omega)
  \alpha_3'' (\omega)\,
  \nonumber\\&\times
		\bigg[
  1 - \dfrac{6 \pi c}{\omega}
  \mathbf{m} \cdot
  \text{Im} \, \mathsf{G}_0 (\Delta \mathbf{X}; \omega)
  \mathbf{m}'
  \bigg].
\end{align}

\section{Angular momentum representation for small particles}\label{app:B}

For exceptionally small particles the rotational state may involve only  a limited set of angular momentum states. In this case, it can be more convenient to describe the dynamics in the  angular momentum basis. 

To expand the jump operators of the master equation (\ref{eq:MEPPFull}) in angular momentum basis, one may express the rotation tensor operator in terms of the conjugate Wigner D-matrices \cite{Biedenharn1984} as
\begin{align}
    \hat{\mathsf{R}} = \sum\limits_{\ell = 0}^\infty \sum\limits_{m,k = - \ell}^\ell \mathsf{R}_{m k}^\ell D_{m k}^{\ell *} (\hat{\mathsf{R}})
\end{align}
with $\mathsf{R}_{m k}^{\ell} ={}
  {(2 \ell +1)}
  \int_{\mathrm{SO}(3)} \diff \mu(\mathsf{R}) \, \mathsf{R} D_{m k}^\ell (\mathsf{R})
  /8 \pi^2
$,
where $\int_{\mathrm{SO}(3)} \diff \mu (\mathsf{R}) 1 = 8 \pi^2$.
To express the $\mathsf{R}_{m k}^{\ell}$ concisely, we introduce the conjugate spherical basis 
$\mathbf{b}_0 = \mathbf{e}_z$, $ \mathbf{b}_{\pm} = (\mp \mathbf{e}_x + i \mathbf{e}_y)/\sqrt{2} $. Defining
$\bm{\delta}_{j} = \sum_{n\in\{-,0,+\}} \delta_{j,n} \mathbf{b}_n$ one then has
$
\mathsf{R}_{m k}^{\ell} ={}
		\delta_{\ell ,1} \bm{\delta}_m \otimes \bm{\delta}_k^*
$.

By inserting an orientation basis one finds
\begin{align}\label{eq:DlmkMatEl}
        \braket{\ell m k | D_{m'' k''}^{\ell'' *} (\hat{\mathsf{R}}) | \ell' m' k'} & = (-1)^{m - k}\sqrt{(2 \ell + 1)(2 \ell'+1)} \notag \\
   & \times \! \!
		\begin{pmatrix}
			\ell & \ell' & \ell'' \\
			- m \! & m' \!& m'' \!
		\end{pmatrix}
  \!
		\begin{pmatrix}
			\ell & \ell' & \ell'' \\
			- k & k' & k''
		\end{pmatrix}
\end{align}
in terms of Wigner 3-$j$ symbols \cite{Edmonds1996}, where we used that the conjugate Wigner-D matrices are the angular momentum eigenstates in orientation representation \cite{Biedenharn1984}.

Due to the selection rules of the Wigner 3-$j$ symbols \cite{Edmonds1996} in Eq.~(\ref{eq:DlmkMatEl}) $\mathsf{R}_{m k}^{\ell \neq 1} = 0$ implies that the angular momentum change due to an emission event is bounded by a single quantum. The same follows for the quantum numbers $m, k$. 

The angular momentum representation of (\ref{eq:MEPPFull}) is obtained by inserting the  Wigner D-matrix expansion of the rotation tensors, as well as angular momentum resolutions of the identity around them, and using Eq.~(\ref{eq:DlmkMatEl}).

In the case of the \emph{linear rotor} the Wigner D-matrices reduce to the spherical harmonics,
$D^{\ell *}_{m 0} (\mathsf{R}) = \sqrt{4 \pi /(2 \ell + 1)} Y_\ell^m (\mathsf{R} \mathbf{e}_3) $
\cite{Biedenharn1984}.
For convenience, we provide  the explicit angular momentum representation of the body-fixed symmetry axis operator $\hat{\mathbf{m}}$ appearing in (\ref{eq:MEPPFullLinRot}). Using recursive relations of the Wigner 3-$j$ symbols \cite{Edmonds1996} its components $\hat{m}_j = \mathbf{b}_j^* \cdot \hat{\mathbf{m}}$ reduce to
\begin{align}
    \hat{m}_{\pm} ={}&
  \sum\limits_{\ell=0}^\infty \sum\limits_{m=-\ell}^{\ell}
		\dfrac{1}{\sqrt{2 (2\ell +1)(2\ell +3)}}
  \\&\times
		\Big[
		\sqrt{(\ell \pm m +1)(\ell \pm m +2)} 
		\ket{\ell +1, m \pm 1} \bra{\ell, m}
		\nonumber\\&- 
		\sqrt{(\ell \mp m +1)(\ell \mp m +2)}
		\ket{\ell m} \bra{\ell+1,m\mp1}
		\Big]\nonumber
  \\
  \hat{m}_0 ={}&
  \sum\limits_{\ell=0}^\infty \sum\limits_{m=-\ell}^{\ell}
  \left[\dfrac{(\ell +1)^2 - m^2}{(2\ell +1)(2 \ell +3)}\right]^{1/2}
  \nonumber\\&\times
		\left[
		\ket{\ell, m} \bra{\ell +1, m}
		+ \ket{\ell +1, m} \bra{\ell, m}
		\right].
\end{align}

\section{Motional heating rates for small particles}\label{app:C}
The heating rates (\ref{eq:PHPPRot}) can be obtained for example
by computing the time derivative of the second moments of the linear and angular momentum,
and noting that the inertia tensor of an ellipsoid is diagonal in the principal axes basis with corresponding eigenvalues $I_1 = m (\ell_2^2 + \ell_3^2)/5 $ and cyclic permutations.
The energy gain associated with linear momentum diffusion along the $j$-th principal axis can be written as
\begin{equation}\label{eq:PHPPCom}
    h^j_{\mathrm{cm}} ={} \partial_t \dfrac{\braket{P_j^2}}{2m}
    ={}
    \dfrac{\hbar^2}{2 m}
    \int\limits_0^\infty \diff \omega \,
    \dfrac{\omega^5 \overline{n}(\omega)}{15 \pi^2 c^5 \epsilon_0}
    (2 \TR [\upalpha'' (\omega)] - \alpha_j''(\omega))
    ,
\end{equation}
while that for the rotation around the $j$-th axis is given by Eq.~(\ref{eq:PHPPRot}).
The factors $2 m \overline{l}^2 / 5 I_i $ appearing in the caption of Fig.~(\ref{fig:Heating}) compensate the dependence of the moments of inertia on the ratio of 
principal diameters.

We note that for materials where $\alpha'' (\omega)$ is strongly peaked around thermal wavelengths, as is the case for silica, one can obtain good results by approximating it 
by sum of delta contributions, as the remaining integrand varies comparatively slowly for all relevant temperatures.

\section{Large-particle limit}
\label{chap:LPlim}

Let us finally discuss how the master equation (\ref{eq:MEGen}) simplifies in the case of large bodies. Specifically, we take the particle extension and the radii of curvature of its (convex) surface to be much greater than the characteristic wave length and the absorption length of the radiation, and we assume the dielectric function to be spatially homogeneous for simplicity.
For sufficiently well-oriented particles we then find that the resulting master equation (\ref{eq:LP}) is fully characterized  by the Fresnel coefficients associated with \emph{incident} radiation.

For large particles contributions from current fluctuations  deep inside the body are damped away long before they reach the surface, and the latter can be considered locally flat. The volume integral in the master equation (\ref{eq:MEGen}) may then be approximated by a surface integral and an additional integration into the particle interior perpendicular to the surface.
At the same time, the Green tensor of the full particle may be replaced by the one of dielectric half space \cite{Buhmann2012}, which still involves a two-dimensional integration over wavevector components.  Using the spatial limit representation of the Lindblad operators (\ref{eq:DefVecA}) and~(\ref{eq:DefVecL}) the latter can be carried out  in the stationary phase approximation. This  yields the master equation
\begin{align}\label{eq:bigone}
	\mathcal{D} \rho
	={}&
	\int\limits_{0}^{\infty} \diff \omega \int\limits_{S^2} \Diff{2} n \int\limits_{\partial V} \Diff{2} r_\mathrm{s} \int\limits_{-\infty}^{0} \diff s^{\perp}
	\dfrac{2 \omega^3}{ \pi c^3}
    \overline{n} (\mathbf{r}_\mathrm{s} + \mathbf{e}^\perp s^\perp, \omega )
    \nonumber\\ & \times
    \text{Im}[\epsilon_\mathrm{r} (\omega)]
	 \TR\bigg[
	\hat{\mathsf{L}} (\mathbf{n}, s^{\perp}, \omega; \mathbf{r}_\mathrm{s})
        \, \rho \,
        \hat{\mathsf{L}}^\dagger (\mathbf{n}, s^{\perp}, \omega; \mathbf{r}_\mathrm{s})
        \nonumber\\&\phantom{R R}
        - \dfrac{1}{2} \left\{ \hat{\mathsf{L}}^\dagger (\mathbf{n}, s^{\perp}, \omega; \mathbf{r}_\mathrm{s})
        \hat{\mathsf{L}} (\mathbf{n}, s^{\perp}, \omega; \mathbf{r}_\mathrm{s})
        , \rho \right\}
	\bigg]
\end{align}
where the Lindblad operators can now be given explicitly, 
\begin{align}
&\hat{\mathsf{L}}_{} (\mathbf{n}, s^{\perp}, \omega; \mathbf{r}_\mathrm{s})
    ={}
    \nonumber\\&
	\dfrac{k}{4 \pi} \sum_{\sigma \in \{s,p\}} \!
	\bigg[
	\mathbf{e}^{\perp} \cdot \hat{\mathsf{R}}^T \mathbf{n} \,
	\Theta (\mathbf{e}^{\perp} \cdot \hat{\mathsf{R}}^T \mathbf{n})
	\nonumber\\&\times
	e^{- i k \mathbf{n} \cdot \hat{\mathbf{X}} }\,
        \dfrac{e^{- i k \mathbf{r}_\mathrm{s} \cdot \hat{\mathsf{R}}^T \mathbf{n} - i k^{\perp}_\mathrm{in} ( \hat{\mathsf{R}}^T \mathbf{n}; \mathbf{e}^{\perp}) s^{\perp}}}
        {k^{\perp}_\mathrm{in} ( \hat{\mathsf{R}}^T \mathbf{n}; \mathbf{e}^{\perp})}
	\nonumber\\&\times
	t_{\sigma} ( \hat{\mathsf{R}}^T \mathbf{n}; \mathbf{e}^{\perp})
    \,
    \mathbf{e}_{\sigma}^\mathrm{ex} ( \mathbf{n}; \hat{\mathsf{R}} \mathbf{e}^{\perp})
    \otimes \mathbf{e}_{\sigma}^\mathrm{in} ( \hat{\mathsf{R}}^T \mathbf{n}; \mathbf{e}^{\perp})
	\bigg]
 .
\end{align}
Here, we introduced the vacuum wavenumber $k=\omega/c$, suppressed frequency dependencies as well as the dependence of the local surface normal $\mathbf{e}^\perp (\mathbf{r}_\mathrm{s})$ on the surface point, and use the label  $m\in\{\mathrm{in}, \mathrm{ex}\}$ to distinguish optical quantities in the interior and exterior of  the body. Defining the normal component of the wavevector by
\begin{equation}
k^{\perp}_{m} (\mathbf{n}; \mathbf{e}^{\perp}) ={}
k \sqrt{ \epsilon_\mathrm{r}^{m} - |\mathbf{e}^{\perp} \times \mathbf{n}|^2},
\end{equation}
with $\epsilon_\mathrm{r}^\mathrm{in} = \epsilon_\mathrm{r}$ ,  $\epsilon_\mathrm{r}^\mathrm{ex} = 1$,  the local polarization vectors read as
\begin{align}
	\mathbf{e}_\mathrm{s}^{m} ( \mathbf{n}; \mathbf{e}^{\perp}) ={}& \dfrac{ \mathbf{n} \times \mathbf{e}^{\perp} }{|\mathbf{n} \times \mathbf{e}^{\perp} |}
	\\
	\mathbf{e}_\mathrm{p}^{m} ( \mathbf{n}; \mathbf{e}^{\perp}) ={}&
	\dfrac{1}{\sqrt{\epsilon_\mathrm{r}^{m}} }
 \Big( |\mathbf{e}^{\perp} \times \mathbf{n}| \mathbf{e}^{\perp}
 \nonumber\\&
	- \dfrac{ k^{\perp}_{m} (\mathbf{n}; \mathbf{e}^{\perp}) }{k} 
 \dfrac{[\mathds{1} - \mathbf{e}^{\perp} \otimes \mathbf{e}^{\perp}]\mathbf{n}
 }{|\mathbf{e}^{\perp} \times \mathbf{n}|}
	\Big),
 \end{align}
where we here use the convention $\mathbf{e}_\mathrm{p}^{m*}\cdot\mathbf{e}_\mathrm{p}^{m} = (k^2 |\mathbf{e}^{\perp} \times \mathbf{n}|^2 + | k_m^\perp (\mathbf{n}, \mathbf{e}^\perp) |^2)/k^2|\epsilon_\mathrm{r}^m| \neq 1$. The local Fresnel coefficients are given by
\begin{align}
	t_\mathrm{s} ( \mathbf{n}; \mathbf{e}^{\perp}) ={}& \dfrac{ 2 k^{\perp}_{\mathrm{in}} (\mathbf{n}; \mathbf{e}^{\perp}) }{k^{\perp}_{\mathrm{in}} (\mathbf{n}; \mathbf{e}^{\perp}) + k^{\perp}_\mathrm{ex} (\mathbf{n}; \mathbf{e}^{\perp})}
	\\
	t_\mathrm{p} ( \mathbf{n}; \mathbf{e}^{\perp}) ={}&
	\dfrac{1}{\sqrt{\epsilon_\mathrm{r}^\mathrm{in}}} \dfrac{2 \epsilon_\mathrm{r}^\mathrm{in} k^{\perp}_{\mathrm{in}} (\mathbf{n}; \mathbf{e}^{\perp})}{k^{\perp}_{\mathrm{in}} (\mathbf{n}; \mathbf{e}^{\perp}) + \epsilon_\mathrm{r}^\mathrm{in} k^{\perp}_\mathrm{ex} (\mathbf{n}; \mathbf{e}^{\perp})}
 .
\end{align}

The master equation (\ref{eq:bigone}) simplifies if the body can be taken well localized in a certain orientation $\mathsf{R}_0$, i.e.\ $\langle\mathsf{R}|\rho|\mathsf{R}'\rangle \neq 0$
only for $||\mathsf{R}^{(\prime)}-\mathsf{R}_0||\ll 1$,
and if the temperature is approximately constant over the depth of the emitting surface layer; this allows computing the $s^\perp$-integral. Remarkably, the remaining factor
\begin{align}
	\nonumber& \! \sum\limits_{\sigma \in \{\mathrm{s}, \mathrm{p}\}}
	\left| \dfrac{t_{\sigma} ( \mathsf{R}^T \mathbf{n}; \mathbf{e}^{\perp})}{ k^{\perp}_\mathrm{in} ( \mathsf{R}^T \mathbf{n}; \mathbf{e}^{\perp})}  \right|^2
	\mathbf{e}_{\sigma +}^\mathrm{in} ( \mathsf{R}^T \mathbf{n}; \mathbf{e}^{\perp}) \cdot \mathbf{e}_{\sigma +}^{\mathrm{in}*} ( \mathsf{R}^T \mathbf{n}; \mathbf{e}^{\perp})
	\\={}&
	\dfrac{1- R_\mathrm{s} ( \mathsf{R}^T \mathbf{n}; \mathbf{e}^{\perp})  + 1 - R_\mathrm{p} ( \mathsf{R}^T \mathbf{n}; \mathbf{e}^{\perp})
 }{
 k^{\perp}_\mathrm{ex} ( \mathsf{R}^T \mathbf{n}; \mathbf{e}^{\perp}) \text{Re} \, k_\mathrm{in}^{\perp} ( \mathsf{R}^T \mathbf{n}; \mathbf{e}^{\perp}) }
\end{align}
can be given in terms of the Fresnel intensity reflection coefficients for \textit{incident} radiation
\begin{align}
	R_\mathrm{s} ( \mathbf{n}; \mathbf{e}^{\perp}) ={} &
	\bigg | \dfrac{k^{\perp}_\mathrm{in} ( \mathbf{n}; \mathbf{e}^{\perp}) - k^{\perp}_\mathrm{ex}( \mathbf{n}; \mathbf{e}^{\perp}) }{k^{\perp}_\mathrm{in} ( \mathbf{n}; \mathbf{e}^{\perp}) + k^{\perp}_\mathrm{ex} ( \mathbf{n}; \mathbf{e}^{\perp}) } \bigg |^2
	\\
	R_\mathrm{p} ( \mathbf{n}; \mathbf{e}^{\perp}) ={} &
	\bigg | \dfrac{k^{\perp}_\mathrm{in} ( \mathbf{n}; \mathbf{e}^{\perp}) - \epsilon_\mathrm{r}^\mathrm{in} k^{\perp}_\mathrm{ex} ( \mathbf{n}; \mathbf{e}^{\perp}) }{k^{\perp}_\mathrm{in} ( \mathbf{n}; \mathbf{e}^{\perp}) +\epsilon_\mathrm{r}^\mathrm{in} k^{\perp}_\mathrm{ex} ( \mathbf{n}; \mathbf{e}^{\perp}) } \bigg |^2
 .
\end{align}
Introducing the transmission coefficients $T_\sigma= 1 - R_\sigma$ and choosing $\mathsf{R}_0 = \mathds{1}$ then gives Eq.~(\ref{eq:LP}).
We remark that Eq.~(\ref{eq:Phis}) can also be obtained directly from (\ref{eq:PhFluxGen}) in an analogous fashion.

The master equation (\ref{eq:LP}) yields heating rates for motion  along and rotations around the body fixed axes $ \mathsf{R}_0 \mathbf{e}_j$, as in the case of small particles, cf. Eqs. (\ref{eq:PHPPCom}), (\ref{eq:PHPPRot}) and Fig.~\ref{fig:Heating}.
As one might expect, they are determined by the second moments of the outgoing linear and angular photon momenta,
\begin{align}
		h^j_\mathrm{cm} ={}&
		\dfrac{\hbar^2}{2 m}
		\int\limits_{0}^{\infty} \diff \omega \int\limits_{S^2} \Diff 2 n \int\limits_{\partial V} \Diff 2 r_\mathrm{s} \,
  \Phi_\mathrm{s} (\mathbf{n}, \mathbf{r}_\mathrm{s}; \omega)
		\dfrac{\omega^2}{c^2} [\mathbf{e}_j \cdot \mathbf{n}]^2
		\\
		h^j_\mathrm{rot} ={}&
		\dfrac{\hbar^2}{2 I_{j}} 
		\int\limits_{0}^{\infty} \diff \omega \int\limits_{S^2} \Diff 2 n \int\limits_{\partial V} \Diff 2 r_\mathrm{s} \,
  \Phi_\mathrm{s} (\mathbf{n}, \mathbf{r}_\mathrm{s}; \omega)
		\dfrac{\omega^2}{c^2} 
  [
  \mathbf{e}_j \cdot (\mathbf{r}_\mathrm{s}  \times\mathbf{n}) ]^2
		,
\end{align}
as characterized by the spectral photon intensity per surface area (\ref{eq:Phis}).

\end{document}